\documentclass[amsmath,amssymb,aps,pra,twocolumn,longbibliography]{revtex4-1}
\PassOptionsToPackage{breaklinks}{hyperref}
\PassOptionsToPackage{breaklinks}{hyperref}
\AtBeginDocument{\RequirePackage{hyperref}}
\usepackage{tikz}

\usepackage{hyperref}
\usepackage{graphicx}
\usepackage{verbatim}
\usepackage[utf8]{inputenc}

\newcommand{\vp}{\varphi}

\newcommand{\vt}{\vartheta}
\newcommand{\tl}{\tilde}

\begin{document}
\title{Can the zero-point energy of the quantized harmonic oscillator be lower? \\  Possible implications for the physics of ``dark energy'' and ``dark matter''}
\author{H. A. Kastrup}
\email{hans.kastrup@desy.de}
\affiliation{DESY Hamburg, Theory Group, Notkestrasse 85, D-22607 Hamburg, Germany}
\begin{abstract}  Replacing the canonical pair $q$ and $p$ of the classical harmonic oscillator (HO) by the \emph{locally and symplectically} equivalent pair
  {\it angle}
  $\vp$  and {\it action variable} $I$ implies a qualitative change of the {global} topological structure of the associated phase spaces: the pair $(q,p)$ is an element of a topologically
  trivial plane $\mathbb{R}^2$ whereas the pair $(\vp,I>0) \in \mathbb{S}^1\times \mathbb{R}_+ $ is an element of a topologically non-trivial, infinitely connected,
  $punctured$ plane $\mathbb{R}^2 -\{0\}$, which has the orthochronous ``Lorentz'' group $SO^{\uparrow}(1,2)$ (or its two-fold covering, the  symplectic group
  $Sp(2,\mathbb{R})$) as its ``canonical'' group. Due to its infinitely many covering groups the resulting  (``symplectic'')
  spectrum of  the associated quantum
  Hamiltonian $H= \omega\, \hat{I}$ is given by $\{\hbar\omega(n+b), n=0,1,\ldots;\, b \in (0.1],\text{e.g.\ $b = 1/s,\, s \in \mathbb{N}$ and large}\}$, in contrast
  to the $(q,p)$ version, where the Hamiltonian has the ``orthodox'' spectrum
   $\{\hbar \omega (n+1/2)\}$.  The deeper reason for the difference is that for  the description of the periodic orbit
   $\{p=p(q)\}$  one covering of $\mathbb{S}^1$ suffices, whereas one generally needs many coverings for the time evolution $\vp(t)$. And this, in turn, can lead to
   a lowering of the zero-point energies.
  
  Several theoretical and possible experimental implications of the ``symplectic'' spectra of the HO are discussed: The potentially most important ones concern the 
  vibrations of diatomic molecules in the infrared, e.g.\ those of molecular hydrogen H$_2$. Those symplectic spectra of the HO may provide a simultaneous key to two
  outstanding astrophysical puzzles, namely the nature of dark (vacuum) energy and that of dark matter: To the former because the zero-point energy $b\,\hbar \omega$ of free electromagnetic wave oscillator modes
  can be extremely small $> 0$ $(b \approx \exp{(-35)}$ for the measured dark energy density). And a key to the dark matter problem because the quantum zero-point energies of the electronic Born-Oppenheimer potentials in which the two nuclei of H$_2$ or the nuclei of other primordial diatomic molecules vibrate can be lower, too, and, therefore, may lead to spectrally detuned ``dark'' H$_2$ molecules during the ``Dark Ages'' of the universe and forming WIMPs in the hypothesized sense! All results appear to be in surprisingly good agreement with the $\Lambda$CDM model of the universe.

  Besides laboratory experiments the search for 21-cm radio signals from the Dark Ages of the universe and other astrophysical observations can help to explore
  those hypothetical implications. \end{abstract} \maketitle \tableofcontents
\section{Introduction}
It very probably appears presumptuous and provocative to question the well-known quantum properties of the   primeval prototype of  quantum mechanical systems: the harmonic oscillator    (HO in the following)!

The motive for daring  a new look at the physical system HO arise from its well-known {\it locally - but not globally -} equivalent canonical descriptions: either in terms of the Cartesian coordinates $(q,p) \in \mathbb{R}^2$ {\it or} in terms of the angle and action variables $(\vp,I) \in \{\mathbb{R} \bmod{2\pi} \times
\mathbb{R}^+\} \cong \mathbb{S}^1 \times \mathbb{R}^+ \cong \mathbb{R}^2 -\{0\}$, where $x \in \mathbb{R}^+$ iff $x \in \mathbb{R}$ and $x>0$, the relationship of which
 can be defined by \cite{kas} 
 \begin{equation} \label{eq:1}
   q(\vp,I) = \sqrt{\frac{2\,I}{m\,\omega}}\, \cos \vp\,, ~~ p(\vp,I) = - \sqrt{2\,m\,\omega\,I}\,
 \sin \vp\,.
   \end{equation}
This mapping is {\it locally} symplectic:
\begin{equation} \label{eq:2}
   dq\wedge dp=d\vp\wedge dI \,,~~\text{or}~~ \frac{\partial(q,p)}{\partial(\vp,I)}=1
   \,.
\end{equation}
 \\ The canonically equivalent Hamiltonians are  given by \begin{equation} \label{eq:3}
  H(q,p) = \frac{1}{2m}\,p^2 + \frac{1}{2}\,m\,\omega^2\,q^2 = H(\vp,I)=
  \omega\,I\,, \end{equation}
with their respective canonical Eqs.\ of motion
\begin{align} \label{eq:4}  \dot{q} = &\frac{\partial H}{\partial p} = p/m,~~~ \dot{p} = -\frac{\partial H}{\partial q}= -m\omega^2q\, ; \\
  \dot{\vp} = &\frac{\partial H}{\partial I}=  \omega,
~~~\dot{I} = -\frac{\partial H}{\partial \vp} = 0, \label{eq:5}
\end{align}
the latter with the obvious solutions
\begin{equation} \label{eq:6}
  \vp(t) = \omega\,t +\vp_0\,,~~I = \text{ const.} = I_0 = E/\omega  > 0\,. \end{equation}
The solutions of the Eqs. \eqref{eq:4} and \eqref{eq:5} describe - as functions of time $t$ - orbits in the repective phase spaces
 \begin{eqnarray}
  \mathcal{S}_{q,p} &=& \{(q,p) \in \mathbb{R}^2\,\} \cong
  \mathbb{R}^2\, \label{eq:7}\\ & & \nonumber \\ \text{ and }~~
  \mathcal{S}_{\vp,I} &=& \{(\vp,I), \vp \in
 \mathbb{R} \bmod {2\pi}\,,\, I > 0\,\} \label{eq:8} \\ & & \cong S^1 \times \mathbb{R}^+ \cong \mathbb{R}^2
 -\{0\}\,. \nonumber \end{eqnarray}

The crucial point - for all what follows in the present paper - is this: the two phase spaces \eqref{eq:7} and \eqref{eq:8} are {\it globally (topologically)}
qualitatively different! Whereas the  $(q,p)$-space is a simply connected and topologically trivial plane $\mathbb{R}^2$, the  $(\vp,I)$-space is topologically
a ``punctured'' plane, i.e. a $\mathbb{R}^2$ with the origin $\{(0,0)\}$ deleted!

This is so for several reasons: the variables $\vp$ and $I$ can be considered as polar coordinates of a plane, with - obviously - $\vp$ the angle and $I>0$ the
radial variable. The value $I=0$ has to be excluded because otherwise the angle
$\vp$ becomes undefined at that point.
In addition the value $I=0$ describes a branch point for the transformation functions \eqref{eq:1}. More arguments can be found in Ref.\ \cite{ka8}.

The topology of the phase space \eqref{eq:8} may equivalently be characterized as that of a simple cone with the tip deleted or as that of a semi-cylinder without the points of the finite circular surface at $I=0$.

If $\vp(t)$ and $I(t)$ describe the moving  points of a periodic orbit on  $\mathcal{S}_{\vp,I}$ then those points may loop around the origin arbitrarily
many times (the first homotopy group $\pi_1$ of $\mathcal{S}_{\vp,I}$ consists of the integers $ \in \mathbb{Z}$), because the orbit coordinate $\vp(t)$ can circle the origin of
$\mathbb{S}^1$ arbitrarily many times in the course of time $t$! Thus, the configuration space of $\vp$ corresponds to one of the infinitely many covering spaces
of the circle $\mathbb{S}^1$, the universal covering being the real line $\mathbb{R}$. A physical example for a high number of coverings is provided by the oscillations of electromagnetic vibrations.

That missing point in the phase space $\mathbb{R}^2_0 \equiv \mathcal{S}_{q,p}-\{0,0\}$, or in $\mathcal{S}_{\vp,I}$, has dramatic consequences for the associated quantum theory
which will be discussed in more detail below.

The crucial result is the following:

The quantum operator version $\hat{I} = K_0$ of the classical action variable $I$ of the HO has the possible spectrum 
\begin{equation}
  \label{eq:31}
  K_0|b,n\rangle = \hbar(n+b)|b,n\rangle,\,\, n=0,1,2,\ldots;\ 1>b> 0.
\end{equation}
Here $K_0$ is the self-adjoint Lie algebra generator of the compact subgroup O(2)  in an irreducible unitary representation of the 3-dimensionsl ``orthochronous
Lorentz'' group $SO^{\uparrow}(1,2) \cong
Sp(2,\mathbb{R})/Z_2$ or of one of its infinitely many covering groups, where the double covering $Sp(2,\mathbb{R})$ is the symplectic group of the plane. (see below). 
This means that the HO Hamilton operator
\begin{equation}
  \label{eq:32}
  H_{\vp,I} = \omega\,K_0,
\end{equation}
associated with the phase space $\mathcal{S}_{\vp,I}$, can have a ground state ($n=0$) with eigenvalue (zero-point energy) $\hbar\omega\, b,\,1 > b>0$, especially with $1/2 > b>0$\;!

As indicated above, the mathematical origin of this possibility is a group theoretical one: The ``canonical'' transformation group of the punctured plane
$\mathbb{R}^2_0 $ is the symplectic group $Sp(2,\mathbb{R})$ which acts transitively on $ \mathbb{R}^2_0 $ (for any two points on
$\mathbb{R}^2_0$ there is an element of $Sp(2,\mathbb{R})$ which connects the two but \emph{leaves the origin invariant})!

The group $Sp(2,\mathbb{R})$ is a twofold covering of the ``orthochronous Lorentz'' group  $SO^{\uparrow}(1,2)$ in ``one time and two space dimensions'' which acts correspondingly on  the phase space $\mathcal{S}_{\vp,I}$, i.e.\ acting transitively and  leaving the origin invariant!    The number $b, \, 1>b>0$  in Eq.\ \eqref{eq:31} characterizes an irreducible unitary representation of a given covering group of $SO^{\uparrow}(1,2)$ \cite{kast}.

The main mathematical aspects of the present paper have  been presented previously in Refs.\ \cite{ka0,boj,ka7,ka8,boj2}. A brief - introductory but probably helpful - summary of them is given in
Ref.\ \cite{ka9}. Essential mathematical references are \cite{barg,puka,vile,sall1,sall2,wolf}.

The present paper tries to draw attention to possible experimental and observational implications of the $(\vp,I)$-framework for the HO. Hopefully, appropriate laboratory experiments and astrophysical obervations will be able to find out whether nature has ``made use of the available mathematical possibilities''(Dirac) or not!

There are - at least - two immediate crucial questions:

i) Why should the classical canonical pair $(\vp,I)$ be a ``better'' - or at least equivalent - basis for the quantum description of a system like the HO, compared to the conventional pair
$(q,p)$?

ii) If the predictions of the quantized $(\vp,I)$ framework are richer  than the $(q,p)$ framework - but not contradictory -, why haven't they been observed yet?

Ad i): A crucial obstacle for using the ``observable'' angle $\vp$ itself for the quantum description of a physical system has been that there exist no corresponding self-adjoint
operator $\hat{\vp}$ \cite{ka,ka2}! This shortcoming can, however, be remedied by the following observation \cite{ka,ka2}:

Geometrically an angle $\vp$ can be defined by
two oriented rays (vectors) both originating from the same given point. The two rays then span a plane. In order to describe the angle uniquely analytically one chooses a third
ray which  ``emanates'' from the same point and which is orthogonal to one of the two original rays. Projecting the second original ray onto the two orthogonal
ones by means of a circle, with radius $a$, around the origin yields a pair $a(\cos\vp, \sin\vp),\, a >0,$ which determines $\vp$ uniquely, after chosing a clockwise- or counter-clockwise orientation.  It is convenient, but not necessary, to put
$a=1$.

Quantizing the system then means: quantizing the components $a\cos\vp$ and $a\sin\vp$ which combined represent one ``observable'', the angle $\vp\,$! The details depend on the choice of $a$ and possibly other elements
of the associated Poisson algebra.

We now come to a crucial physical point:

For periodic motions - like that of a HO - the angle $\vp(t)$ of  Eq.\ \eqref{eq:5} does not stop at $\vp(T) \equiv \omega\,T = 2\pi$ but ``runs'' around the origin,
say at least $s $ times, i.e.\ $\vp(t) \geq \vp(t_s) = s\,2\pi,\, s \in \mathbb{Z},$, and in this way generates an (s+1)-fold covering of the unit circle $\mathbb{S}^1$. In this way the configuration
space $\mathbb{S}^1_{[s+1]}$ of the angle becomes (s+1)-fold connected, and, as $s$ can be an arbitrary integer, infinitely connected.

So, for dynamical (time-dependent) periodic systems the ``observable'' angle consists of 2 parts: the number $s$ of completed coverings  $\mathbb{S}^1_{[s]}$ of the unit circle and a ``rest'' $\chi \in [0,2\pi)$:
\begin{equation}
  \label{eq:266}
  \vp = (2\pi\,s + \chi) \in \mathbb{S}^1_{[s+1]}, ~ s \in \mathbb{Z},~ \chi \in [0,2\pi).
\end{equation}
The crucial point for periodic motions is that $s=0$ is sufficient  to describe the orbit $p(q)$ of Eq.\ \eqref{eq:1}, but in order to describe the time evolution $\vp(t)$ one needs
to know the pair $(\chi;s)$ of Eq.\ \eqref{eq:266}. This is a consequence of the non-trivial topology of the phase space \eqref{eq:8}. In many cases the angle $\vp(t)$
appears in the form $\vp(t) = \omega t \equiv \tl{t} $, i.e. it is essentially a time variable. Thus, the pair $(q,p) \in \mathcal{S}_{q,p}$ of a periodic orbit is independent of the number $s$ of coverings. But this number of coverings is essential in connection with the phase space $ \mathcal{S}_{\vp,I}$. This important difference
leads to corresponding different quantum mechanical properties of the two phase spaces, e.g.\ for  $\mathcal{S}_{\vp,I}$ to the set of spectra \eqref{eq:31}, containing
the ``orthodox'' case $b=1/2$ as a special one! We shall see that an $s$-fold covering ($s >0$ ) is associated with $b = 1/s$. 

The introduction of the  canonical pair angle and action variables is conventionally motivated by the aim to make the action variable $I$ a constant of motion, i.e.\ to have
an ``integrable'' system \cite{land,thir,arn}. But this is not necessary: one can try to  describe systems in the phase space \eqref{eq:8} in terms of the
local coordinate pair $(\vp,I)$ or the global ones $h_0=I, .\,h_1 = I\cos\vp,\, h_2 -I\sin\vp$   which will be illustrated by an example in the next chapter.

Phases play an important role in many physical systems with periodic properties like  vibrations, waves etc.,  e.g.\ in optics, atomic and molecular spectroscopy, condensed matter physics etc.\  Thus, it is important to understand the corresponding quantum theories in terms of the canonical pair angle and action variable properly and consistently and look for experimental consequences.

Ad ii): One reason might be that nobody up to now has been looking for the newly predicted physical phenomena! Another reason could be that the associated signals are very weak and obscured by the ``orthodox'' spectrum $\hbar \omega \, (n+ 1/2)$!  Possible related future laboratory experiments are indicated
in Ch.\ IV  and associated interpretations of present and future astrophysical observations in Ch.\ V.

The paper is organized as follows:

Ch.\ II discusses a few properties of the phase space \eqref{eq:8}: its global coordinates $h_j, j=0,1,2$,  provided in terms of the group $SO^{\uparrow}(1,2)$. Further  the trivial orbits (circles) of the HO on the phase space \eqref{eq:8} and those of a dynamical model, a simple generalization on  $\mathcal{S}_{\vp,I}$ of the HO. These ``classical'' considerations are intended to provide an intuitive background for the discussions of the associated quantum mechanics in Ch.\ III.

In Ch.\ III several aspects of quantum mechanical systems are discussed the basic ``observables'' of which are given by the Lie algebra elements $K_0, K_1$ and $K_2$ of the canonical group $Sp(2,\mathbb{R})$  of the punctured plane. Though this Lie algebra is also that of the isomorphic groups $SU(1,1)$ or $SL(2, \mathbb{R})$, of the group  $SO^{\uparrow}(1,2)$ and all covering groups as well,  the ``symplectic'' variant appears to be preferable because of  possible generalizations to higher dimensional phase spaces \cite{ka8,ka7}.

One of the main topics in this chapter consists of the discussion of the spectrum
\eqref{eq:31} and related explicit Hilbert spaces for the representation of the self-adjoint operators $K_0, K_1, K_2$. Extended use is made of mathematical results contained in Refs.\ \cite{ka7,ka8}. Hardy spaces (i.e.\ Hilbert spaces which have non-vanishing Fourier components for $n=0,1,2,\ldots$ only) on the circle play a prominent role for constructing those explicit Hilbert spaces.

Ch.\ IV contains a number of suggestions to find concrete physical systems to which the theoretical framework may apply.

Possibly the most important application concerns the vibrations of diatomic molecules which are harmonic in the neighbourhood of the minima of their
Born-Oppenheimer (BO)  potentials and where ``symplectic'' spectra \eqref{eq:31} may lead to a lower ($b < 1/2$) ground state energy compared to the ``orthodox'' value $b=1/2$\;!

Laboratory tests are, of course, of crucial importance, especially for molecular hydrogen H$_2$. As this molecule has no permanent electric dipole moment, its
``orthodox'' infrared emission and absorption signals are already very weak. Therefore very probably even more so the ``non-orthodox'' ones.  Other diatomic molecules of the lightest elements with an electric dipole element (like, e.g.\ LiH) may be more appropriate for laboratory infrared experiments. 

For H$_2$ itself Raman scattering or atomic and molecular collisions  may induce transient electric dipole moments leading to characteristic emissions or absorptions associated with vibrations (and rotations) \cite{herz2}.

Extremely important are possible astrophysical applications, discussed in Ch.\ V,  especially concerning the problems of dark energy and dark matter; here the spectrum \eqref{eq:31}  may
provide the key to the simultaneous understanding of both problems:

As the index $b$
may be arbitrarily small $>0$, the associated estimate of the cosmological constant $\Lambda$ - or the vacuum ``dark'' energy density - can be compatible with the experimentally observed value,
leading to $b \approx \exp(-35)$\:. 

In addition, the possible lowering of the vibrational zero-point energies of electronic Born - Oppenheimer potentials for diatomic molecules suggests to look at molecular hydrogen b-H$_2$ and other primordial diatomic molecules as  candidates for dark matter.

Altogether one finds that the consequences of a symplectic spectrum ($0< b< 1/2$) of the HO are surprisingly well compatible with the cosmological $\Lambda$CDM model, with - mainly - $b$-H$_2$ molecules as WIMPs~!

There is, however, one important caveat: the dynamics of the transitions (rates) to and from the new additional energy levels has still to be worked out\,!

Experimentally, 21-cm radio telescopes directed towards the Dark Ages of the universe are of special importance (see, e.g.\ Ref.\ \cite{koop}). Recent observations \cite{bow} indicate -- unexpected for the present interpretations of dark matter -- non-gravitational (electromagnetic?) interactions between atomic hydrogen and dark matter \cite{bark,fial2}\,! This appears to be compatible with $b$-H$_2$ molecules as dark matter
Similarly the recently observed discrepancy between computer simulated dark matter models and gravitational lensing \cite{mene} is of interest in this context.

If the observed cosmic dark matter indeed consists of - infrared detuned - primordial diatomic molecules then there is no need for the introduction of any kind
of ``new'' matter, a point which has also been emphasized in the recent discussions of dark matter as being formed by primordial black holes (for a recent review see,
e.g.\ Ref.\ \cite{has}).

\section{Motions on the classical \\ phase spaces $\mathcal{S}_{\vp,I}$ and $ \mathcal{S}_{q,p}$}
The present chapter discusses a simple classical model on the phase space $\mathcal{S}_{\vp,I}$ as a preparation for the discussion of the corresponding quantum
mechanical one later.
\subsection{Coordinates and orbits on $\mathcal{S}_{\vp,I}$ }
\subsubsection{Global coordinates}
It was already indicated above that the angle $\vp$ itself is not a ``good'' global coordinate on $\mathcal{S}_{\vp,I}$. The situation is even worse for the corresponding
quantum theory \cite{ka}. As described above, a way out is to characterize the geometrical quantity ``angle'' by the pair $(\cos \vp, \sin \vp)$. However the triple
$I,C(\vp)=\cos\vp, S(\vp)=\sin\vp$ is
still not appropriate for our present purpose:

Consider the Poisson brackets
\begin{equation} \label{eq:9} \{f_1,f_2\}_{\vp,I} \equiv \partial_{\vp}f_1\,\partial_I f_2- \partial_{I}f_1\,\partial_{\vp} f_2 \end{equation}
for locally smooth functions on $\mathcal{S}_{\vp,I}$. The 3 functions $I,C(\vp)$ and $S(\vp)$ obey the Poisson Lie algebra
\begin{equation} \label{eq:10}
 \{I,C\}_{\vp,I}=
S\,,~\{I,S\}_{\vp,I}=-C\,,~ \{C,S\}_{\vp,I}=0\,,
\end{equation}
which constitutes the Lie algebra of the Euclidean group $E(2)$ of the plane: rotations (generated by $I$) and 2 independent translations (generated by $C$ and $S$).
They are the proper coordinates for a phase space with the topology of an {\em infinite} cylinder $\mathbb{S}^1 \times \mathbb{R}$, like that of the canonical system angle and orbital angular
momentum \cite{ka,ka2}.

It can be justified systematically \cite{ka10} that the appropriate global coordinates on  $\mathcal{S}_{\vp,I}$ are the functions
\begin{align} \label{eq:11}
 h_0(\vp,I)& = I >0\,, \\ h_1(\vp,I)=I
\cos \vp\,,&\;\;\; h_2(\vp,I)=-I \sin \vp\,, \nonumber
\end{align} which obey
\begin{equation} \label{eq:228} \vec{h}^2 \equiv h_1^2 + h_2^2 = h_0^2,\,h_0 > 0.\end{equation}
and, therefore, describe a simple (``light'') cone, with the tip deleted.
The functions $h_j(\vp,I)$ obey the Poisson Lie algebra
\begin{align} \label{eq:12}
 \{h_0,h_1\}_{\vp,I}=
-h_2\,,&~\{h_0,h_2\}_{\vp,I}=h_1\,,\\ \{h_1,h_2\}_{\vp,I}&=h_0\,,\nonumber
\end{align}
which constitutes - as mentioned above - the Lie algebra of the 3-dimensional group $SO^{\uparrow}(1,2)$ or of the   symplectic group $Sp(2,\mathbb{R})$ of a $(x,y)$-plane, the transformations of which leave the skew-symmetric form $dx \wedge dy$ invariant.

 The  triple $(h_0,h_1,h_2)$ transforms as a 3-vector with respect to the group $SO^{\uparrow}(1,2)$, the pair $(q,p)$ transforms as
a vector with respect to the symplectic group \cite{ka8}!

As the symplectic group $Sp(2,\mathbb{R})$ is isomorphic to the groups $SL(2,\mathbb{R})$ and $SU(1,1)$ \cite{kas3}, one may use those here, too. But the identification as the symplectic group appears to be more appropriate in the framework of classical mechanics and, above all, it can be generalized to higher dimensions \cite{ka8}.

Justification of the global ``canonical'' coordinates \eqref{eq:11} in a nutshell \cite{ka9}: The three 1-dimensional subgroups of the (transitive) group $SO^{\uparrow}(1,2) = Sp(2,\mathbb{R})/Z_2$ (1 rotation, 2 ``Lorentz boosts'') generate global orbits on  $\mathcal{S}_{\vp,I}$. The generators of these orbits are global Hamiltonian vector fields the associated Hamiltonian functions of which are the ``coordinates'' \eqref{eq:11}. This is in complete analogy to the usual phase space  $\mathcal{S}_{q,p}$ the global coordinates $q$ and $p$ of which are the Hamiltonian functions of the vector fields which generate the global translations in $p$- and $q$-directions on  $\mathcal{S}_{q,p}$, endowed with a symplectic structure in terms of the Poisson bracket $\{\cdot, \cdot\}_{q,p}$.
\subsubsection{Orbits on $\mathcal{S}_{\vp,I}$}
\begin{figure}[h]\includegraphics[width=\columnwidth]{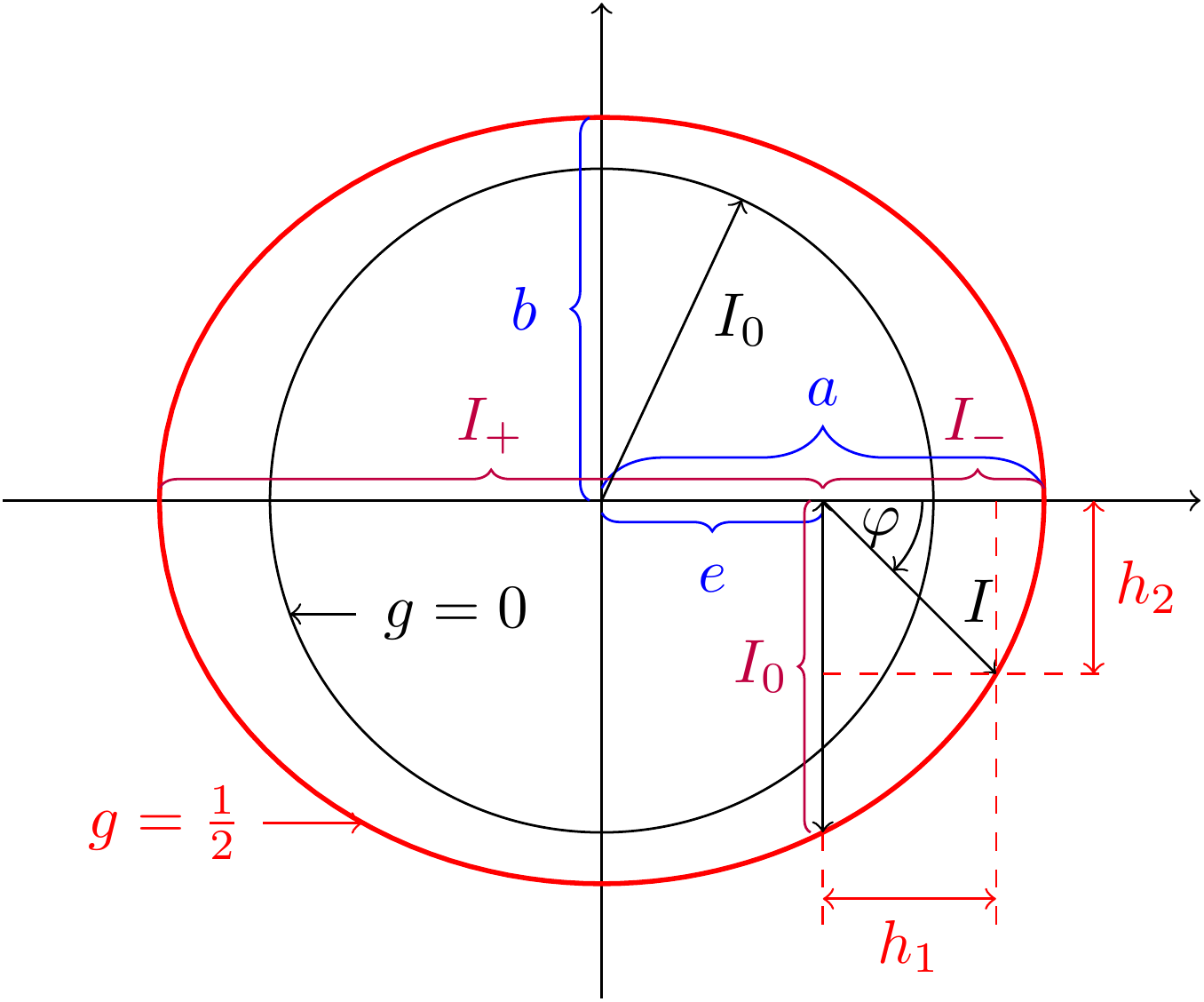}\caption{\label{fig:plot-p-1a} Phase space $ \mathcal{S}_{\vp,I}$: 1.\ Graph of the circular orbit
     $\{(\vp \in [0,2\pi), I=I_0 >0)\}$ generated by the periodic motion \eqref{eq:6}   of the "orthodox"  HO (in black). 2.\ Graph of the   elliptical orbit
     $I(\vp)= I_0/(1+g\,\cos\vp) $    associated with the periodic motions  \eqref{eq:21} and \eqref{eq:286} (in red, with g=1/2). Origin of the polar coordinates $(\vp,I)$ here is the right focus.
     The meaning of the different letters with their related quantities and their historical names are described in the list
     \eqref{eq:267}}
\end{figure}

The  graph of the motion \eqref{eq:6} in $\mathcal{S}_{\vp,I}$ is utterly simple: a circle of radius $I_0> 0$ on which the position at time $t$ is given
by the angle $\vp(t) = \omega t$. We assume that $\vp(t=0)~= 0$ and that $\vp(t)$ starts {\it clockwise} off  a given ray emanating from the point $I=0$.
That ray also defines an horizontal abscissa of an orthogonal coordinate system with an ordinate of pointing upwards (see Fig.\ \ref{fig:plot-p-1a}). The clockwise orientation of the angle 
is induced by the choice of $h_1$ and $h_2$ in Eqs.\ \eqref{eq:11}.

{\it Note that $h_1$ is the projection of $I$ on the positive abscissa and $h_2$ the one on the negative ordinate} (see Fig.\ \ref{fig:plot-p-1a}).

Note also that these two projections do not commute (see the last of the Eqs.\ \eqref{eq:12}), again a consequence of the fact that the point $(0,0)$ does not
belong to the phase space!

Things become more interesting if we ``disturb'' the HO by introducing new interactions. Note that on $\mathcal{S}_{\vp,I}$ functions have to be expressed in terms
of the basic variables $h_j (\vp,I),j=0,1,2$. This means for the HO:
\begin{equation}
  \label{eq:13} H(\vp,I)= \omega\,h_0.
\end{equation}
A simple but interesting modification is \cite{ka11}
\begin{equation}
  \label{eq:14}
  H(\vp,I) = \omega (h_0 + g\,h_1) = \omega\, I(1+ g\,\cos\vp),~ g\ge 0,
\end{equation}
with the Eqs. of motion
\begin{align}
  \label{eq:15}
  \dot{\vp} &= \partial_I H(\vp,I) = \omega(1+g\cos\vp), \\
  \dot{I} &= - \partial_{\vp}H(\vp,I) = g\,\omega I\sin \vp.
            \label{eq:16} 
\end{align}
Now $I(t)$ is no longer a constant.

The present discussion is an extension of the usual one  for (completely) integrable systems \cite{born,land,thir,arn} in which the action variables are constants
of motion as functions on the original $(q,p)$-phase space and where the original ``tori'' (determined by $ I_0=$ const. and $\vp \in \mathbb{S}^1$ in our very special
case) are rather stable against small perturbations (KAM theory \cite{kam}).

Here the {\it global} phase space formed by  angle and action variables is being considered
and the action variable $I$ may be a function of time $t$, like the angle variable $\vp$.

Recall that an action variable is originally defined as a global variable -
like the energy - on the phase space \eqref{eq:7}, namely as a closed path integral along the border of a volume determined by the energy and the potential of the  system \cite{born,land,thir,arn}:
\begin{equation}
  \label{eq:278}
  I(E) = \frac{1}{2\pi}\oint_{C^{\uparrow}(E)}dq\,p(q,E),\,\, p(q,E) = \pm \sqrt{2m(E-V(q)},
\end{equation}
where the clockwise oriented closed path $C^{\uparrow}(E)$ is determined by the energy equation $p^2/2m + V(q) = E$. According to Stokes' theorem the path integral is
equal to the volume with the border $C^{\uparrow}(E)$. Here $I(E)$ is a constant of motion because $E$ is a constant along the orbits $\{[q(t),p(t)]\}$.

The more general case $I(t)$ can be obtained from the local relation \eqref{eq:2}: Integrating both sides simultaneously at time $t$ gives
\begin{equation}
  \label{eq:284}
\Delta  \vp (t)\cdot \Delta I(t) = \int_{\Delta G_{q,p}(t)}dqdp = \Delta V_{q,p}(t),
\end{equation}
where $\Delta V_{q,p}(t)$ is the volume of the region $\Delta G_{q,p}(t) \subset \mathcal{S}_{q,p}$, with $\Delta V_{q,p}(t)=0$ for $\Delta  \vp (t)= 0$ or $ \Delta I = 0$. Putting the lower value $I_0 = 0$ in $\Delta I = I -I_0$ the special case
\eqref{eq:278} is obtained for $\Delta  \vp (t) = 2\pi$.

With $I_0 = 0$ it follows from Eq.\ \eqref{eq:284} that 
\begin{equation}
  \label{eq:285}
  I(t) = \partial V_{q,p}(t)/\partial \vp,
\end{equation}
where the pair $(q,p)$ is assumed to be a function of $\vp$ like in Eqs.\ \eqref{eq:1}. Thus, $I(t)$ may be interpreted as the differential change with $\vp$ of the
phase space volume $V_{q,p}(t)$ at time $t$.

The additional term $g\,h_1$ in the Hamiltonian \eqref{eq:14} breaks several related symmetries: rotation invariance in the $(h_1,h_2)$-plane (which can be remedied
by using the combination $\cos \alpha\, h_1 + \sin \alpha\, h_2$ instead of $h_1$), special
Lorentz ``boosts'' in the directions ``1'' or ``2'' and reflection parity ($\vp \to \vp \pm \pi$). Time reversal ($\vp \to -\vp$) is fulfilled.

The  Eqs.\ of motion \eqref{eq:15} and \eqref{eq:16} can be integrated immediately:

As the energy is still conserved,
\begin{equation}
  \label{eq:17}
  \omega\, I(1+ g\,\cos\vp) = E = \text{$const.$ },
\end{equation}
we have the orbit equation
\begin{equation}
  \label{eq:18}
  I(\vp) = \frac{I_0}{1+g\,\cos\vp},\,\ I_0= E/\omega\,. 
\end{equation}
This equation describes a conical section with a given focus as the origin for the polar coordinates $\vp$ (``true anomaly''), distance $I(\vp)$ from that focus ,
``semi-latus rectum'' $I_0$ and ``numerical eccentricity'' $\epsilon = g$.

For $g<1$ we have an ellipse, for $g=1$ a parabola and for $g>1$ a hyperbola! The angle $\vp$ increases clockwise from the fixed ray which starts from the focus 
 nearest to the orbit point
 $I_- =I_0/(1 +g)$, the ``perihelion'', and further passes through that latter point (see Fig.\ \ref{fig:plot-p-1a}). \pagebreak[2]
 \begin{equation}
   \label{eq:267}
   \text{\bf Ellipse,\; $g< 1: $} 
 \end{equation}
 \begin{itemize} 
 \item semi-latus rectum: $I_0$,
 \item numerical eccentricity: $g$,
 \item perihelion: $I_- = I_0/(1+g)$,
   \item aphelion: $I_+ = I_0/(1-g)$
 \item major semi-axis: \\ $a=\frac{1}{2}[I_0/(1+g) + I_0/(1-g)] = I_0/(1-g^2)$,  
 \item linear eccentricity: $e=g\,a  = gI_0/(1-g^2)$ \\
   (2$e$ is the distance of the two foci),
   \item  minor semi-axis: 
  $ b= \sqrt{a^2 - e^2} = I_0/(1-g^2)^{1/2}$,
\item area of ellipse: $ ab\pi = \frac{I_0^2\pi}{(1-g^2)^{3/2}}$ .
\end{itemize}
Thus the shape of the ellipse is completely determined by the coupling
constant $g$ and the integration constant $I_0$.
 \pagebreak[2] \newline
The same holds for the hyperbola with the focus of the left branch as the origin  for the
polar coordinates:

\begin{equation}
  \label{eq:268}
  \text{\bf Hyperbola,\; $g> 1$:}
\end{equation}
\begin{itemize} \item
  If $g>1$ the expression \eqref{eq:18} describes a hyperbola of which we consider one branch only: the one open to the left.
  Its point of closest distance to the (inside) focus is $I_0/(1+g)$
where $\vp = 0$.

 \item semi-latus rectum: $I_0$,
 \item numerical eccentricity: $g$,
 \item linear eccentricity: $e = gI_0/(g^2-1)$ \\
   (2$e$ is the distance of the 2 foci),
   \item major and minor semi-axis: \\
     $a = I_0/(g^2-1),\,\,\, b= I_0/(g^2-1)^{1/2}$,
   \item The two angles $\vp_{\infty}(i),\,i=1,2$  characterizing the asymptotes are determined by
     $\cos\vp_{\infty}(i) = - 1/g,\,\vp_{\infty}(1) \in (\pi/2,\pi)\text{ and } \vp_{\infty}(2) \in (\pi, \pi + 3\pi/2)$.
   \end{itemize}
   \begin{equation}
     \label{eq:269}
    \text{\bf Parabola,\: $g=1$:} 
   \end{equation}
    
     This simple case can be treated in the same way as the two others above.
   
   Obviously, the orbits of the last two cases extend to infinity.
   \subsubsection{Time evolution}
   
   {\bf Ellipse} \\
The time evolution follows from Eq.\ \eqref{eq:15}:
\begin{equation}
  \label{eq:19}
  \int_{\vp_0}^{\vp}\frac{d\vt}{1+g\,\cos \vt} = \omega\,(t-t_0).
\end{equation}
For $g<1$ wir get \cite{grad1} with $\vp_0=0$ for $t_0=0$:
\begin{equation}
  \label{eq:20}
  \sqrt{1-g^2}\,  \omega\,t = 2\arctan\left[\sqrt{\frac{1-g}{1+g}}\tan(\vp/2)\right], 
\end{equation}
or
\begin{equation}
  \label{eq:21}
\tan[  \vp(t)/2] =  \sqrt{\frac{1+g}{1-g}}\tan(\sqrt{1-g^2}\,\omega\,t/2).  
\end{equation}
Thus, the interaction $g\,h_1(\vp,I)$ leads to an {\it effective redshifted angular frequency}
\begin{equation}
  \label{eq:93}
  \omega_g = \sqrt{1-g^2}\,\omega,
\end{equation}
with a branch point for $g^2 \to 1$.

It follows from Eq.\ \eqref{eq:20} that the time $t(\pi/2)$ needed to pass from $\vp = 0$ to $\vp =\pi/2$ is given by
\begin{equation}
  \label{eq:22}
  \omega_g\,t(\pi/2) = 2\arctan\sqrt{\frac{1-g}{1+g}} = \arccos g.
\end{equation}
At that time $I[t(\pi/2)] = I_0$ (see Eq.\ \eqref{eq:18}).

The time needed to pass from $\vp=0$ to $\vp =\pi$ is
\begin{equation}
  \label{eq:23}
  \omega_g\,t(\pi) = \pi.
\end{equation}
Here we have $I[t(\pi)] =I_0/(1-g)$ (``aphelion'').

For reasons of symmetry of the ellipse we get from Eq.\ \eqref{eq:23} for one period
\begin{equation}
  \label{eq:24}
  \omega_g\,T_{g;2\pi} = 2\pi,\,\, T_{g;2\pi} = \frac{2\pi}{\omega\sqrt{1-g^2}}.
\end{equation}
Thus, $1/\sqrt{1-g^2}$ is a kind of ``refractive index''.

Once the time evolution  $\vp(t)$ is known that of $I(t)$ can be obtained from the orbit equation \eqref{eq:18}. Using the relation $\cos\vp = 1-\tan^2(\vp/2)/
[1+\tan^2(\vp/2)]$ one obtains
\begin{equation}
  \label{eq:286}
  I(t) = \frac{I_0}{1-g^2}(1-g\cos[\omega_gt]),
\end{equation}
which again gives the relations \eqref{eq:22}, \eqref{eq:23} and \eqref{eq:24}.

The above results may be looked at as follows: For vanishing $g$ we have on $\mathcal{S}_{\vp,I}$ a clockwise  periodic motion with frequency $\omega$ on a circle of radius $I_0$.  Adding the interaction $g\,h_1(\vp, I),\,0\leq g <1,$ deformes the circle into an ellipse with semi-latus rectum $I_0$ and numerical
eccentricity $g$. In addition the original angular frequency $\omega$ of the periodic motion is reduced to $\omega_g = \sqrt{1-g^2}\,\omega$.

Pictorially speaking we start with a ``circularly polarized'' motion ($g=0$) which encounters a medium ($0< g<1$) which induces an ``elliptical polarization''
and reduces the original angular frequency $\omega$!

{\bf Hyperbola} \\
As before the time evolution can be calculated from Eq. \eqref{eq:15}, the integration of which now gives \cite{grad1}
\begin{equation}
  \label{eq:25}
  \tan[\vp(t)/2] = \sqrt{\frac{g+1}{g-1}}\tanh[\sqrt{g^2-1}\,\omega\,t/2]. 
\end{equation}

{\bf Parabola}\\
Finally, the time evolution for the parabola ($g=1$) is given by \cite{grad2}
\begin{equation}
  \label{eq:26}
  \tan[\vp(t)/2] = \omega\,t\,.
\end{equation}

\subsection{Orbits on $\mathcal{S}_{q,p}$}
Using the mappings \eqref{eq:1} the orbit equation \eqref{eq:18} in  $\mathcal{S}_{\vp,I}$ can be mapped onto $\mathcal{S}_{q,p}$, where it has the parametrization
\begin{align}
  q_g(\vp) =& \sqrt{\frac{2I_0}{m\omega(1+g\cos\vp)}}\cos\vp \label{eq:107} \\
  p_g(\vp) =& - \sqrt{\frac{2m\omega I_0}{1+g\cos\vp}}\sin\vp.\label{eq:108}
\end{align}
\begin{figure}[h]\includegraphics[width=\columnwidth]{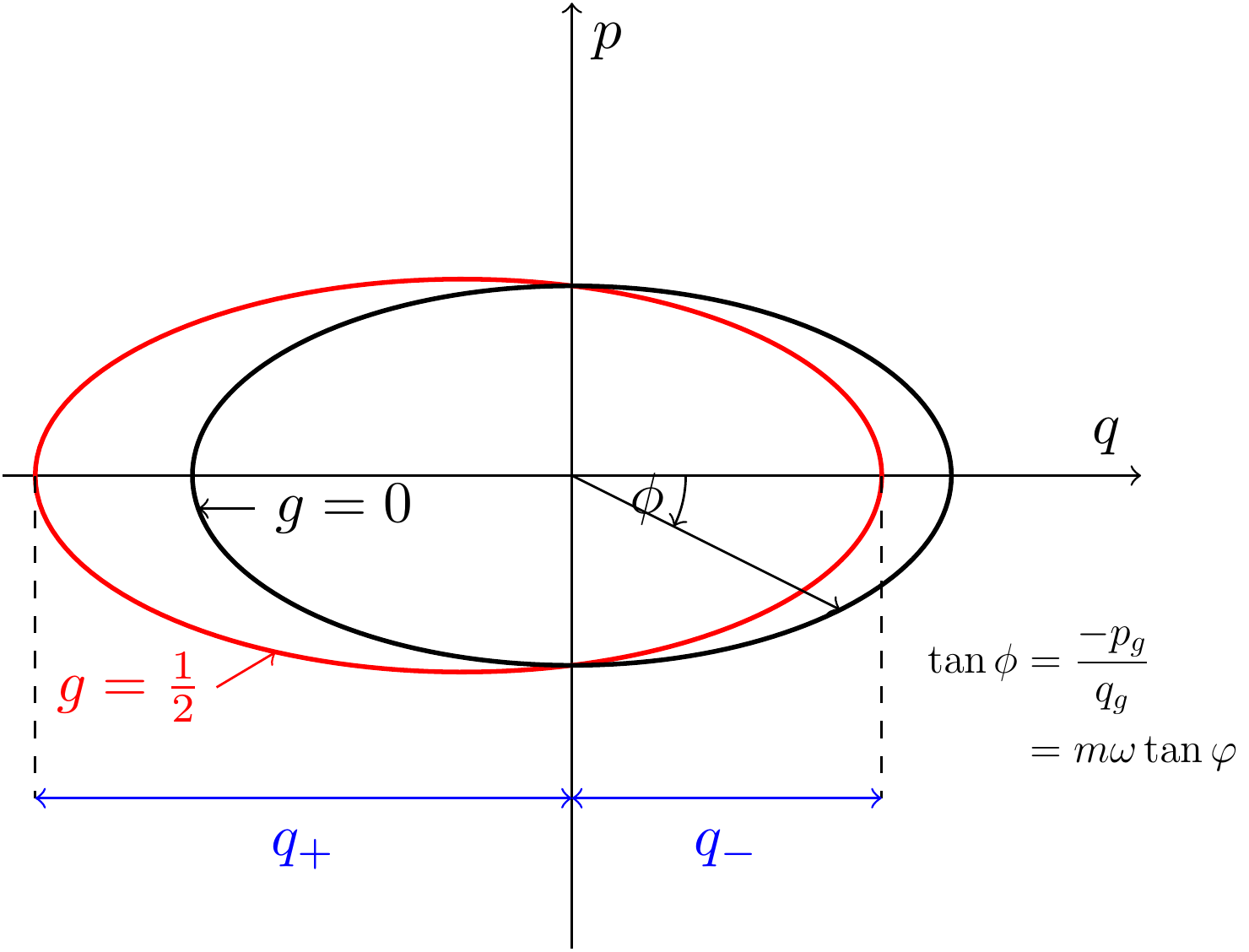}\caption{\label{fig:plot-p-2} Phase space $ \mathcal{S}_{q,p}$: 1.\ Graph of the ``orthodox'' elliptical orbit generated by the usual HO (in black) 2.\ Graph  of the  image of the (red) ellipse in Fig.\ \ref{fig:plot-p-1a} as described by Eqs.\ \eqref{eq:107} and \eqref{eq:108}   (in red). Here $\phi$ is the polar angle on $\mathcal{S}_{q,p}$ and $\vp$ the polar angle on $\mathcal{S}_{\vp,I}$.} \end{figure}
It implies
\begin{align}\label{eq:109}
  H(q_g,p_g) & =\frac{p_g^2}{2m} + \frac{1}{2}m\omega^2\,q_g^2 \\
 & = \omega\,I=\frac{\omega\,I_0}{1+g\,\cos\vp}, \nonumber \\ &
  -\frac{p_g}{q_g} \equiv \tan \phi = m\, \omega \tan \vp. \label{eq:151}
\end{align}
and
\begin{equation}\label{eq:110}
  \cos\vp = \frac{\sqrt{m/2}\;\omega\,q_g}{\sqrt{H(q_g,p_g)}} = \pm \sqrt{V(q_g,p_g)/H(q_g,p_g)}.
\end{equation}
Inserting the last expression into Eq.\ \eqref{eq:109} yields a 4th order equation in $q_g$ and $p_g$ for the orbit described by Eqs. \eqref{eq:107}
and \eqref{eq:108}:
\begin{align}
  \label{eq:27} & H(q_g,p_g) \pm g\,\sqrt{V(q_g,p_g)\,H(q_g,p_g)} = \omega\,I_0, \\
    &[H(q_g,p_g)-\omega\,I_0]^2 = g^2\,V(q_g,p_g)\,H(q_g,p_g).   \nonumber
\end{align}
For $g=0$ this is reduced to the usual orbit ellipse of the HO on $\mathcal{S}_{q,p}$ (See Fig.\ \ref{fig:plot-p-2}):
\begin{equation}
  \label{eq:28}
  \frac{p_{g=0}^2}{2m} +  \frac{1}{2}m\omega^2\,q_{g=0}^2 = E_0 = \omega I_0 = \text{const. }
\end{equation}

The functions \eqref{eq:107} and \eqref{eq:108} have the special values
\begin{align}\label{eq:106}
  q_g(0) \equiv q_- &=  \sqrt{\frac{2I_0}{m\omega(1+g)}},\,\,\, p_g(0)=0,  \\ 
  q_g(\pi/2)&= 0,\,\,~~ p_g(\pi/2) = -\sqrt{2m\omega I_0}, \nonumber \\
  q_g(\pi) \equiv q_+ &= -\sqrt{\frac{2I_0}{m\omega(1-g)}},\,\,\, p_g(\pi)=0, \nonumber \\
            q_g(3\pi/2)&= 0,\,\,~~ p_g(3\pi/2) = \sqrt{2m\omega I_0}. \nonumber
\end{align}
Note that the 4th order figure defined by Eq.\ \eqref{eq:27} is symmetric with respect to the $q$-axis, but no longer symmetric with respect to the $p$-axis
(see Fig.\ \ref{fig:plot-p-2}). 
Maximum and minimum of $p_g(\vp)$ are given by the angles $\vp_{\pm}$, which obey
\begin{equation}
  \label{eq:200}
  \cos \vp_{\pm} =-(1-\sqrt{1-g^2})/g.
\end{equation}
\subsection{The frequency $\omega$ as an external field}
According to Eq. \eqref{eq:3} the Hamiltonian of the HO on $\mathcal{S}_{\vp,I}$ has the simple form
\begin{equation}
  \label{eq:29}
  H(\vp,I) = \omega\,I,
\end{equation}
where the frequency $\omega$ appears as a  parameter multiplying the basic action variable $I$.

 That parameter $\omega$ may also be considered as an external ``field'' which can be ``manipulated'' from outside, e.g., as a function $\omega(t)$ of time $t$.
The solution of Eq.\ \eqref{eq:4} then is
\begin{equation}
  \label{eq:30}
  \vp(t) - \vp(t_0) = \int_{t_0}^td\tau\,\omega(\tau).
\end{equation}
Note that the Hamiltonian \eqref{eq:29} is independent of $\vp$ and therefore $\dot{I} = -\partial_{\vp}H(\vp,I) = 0$, even if $ \omega = \omega (t)$! 
Thus, the action variable $I$ is still conserved on $\mathcal{S}_{\vp,I}$, but the energy $E = \omega (t)\, I$ is not!
A possible interesting example for applications is a time dependent angular frequency of the form
\begin{equation}
  \label{eq:236}
  \omega (t) = \omega_0 + \lambda\sin(\rho t),\; |\lambda| < \omega_0.
\end{equation}
$~~$ In the present context it is also appropriate to briefly recall the so-called ``adiabatic invariance'' \cite{born1,born3,land,thi,arn} of the action variable I: If energy $E(\lambda)$
and frequency $\omega(\lambda)$ of a periodic motion with period $T$ depend on a slowly varying parameter $\lambda$ (~$T\, d\lambda/dt \ll \lambda$), then $I = E(\lambda)/\omega(\lambda)$ remains constant if $\lambda$ varies.

In the following discussions on the ``non-orthodox'' quantum mechanics of the HO it is essential to differentiate between the quantum counterparts of the action
variable $I$ and the Hamilton function $H$, the generator of time evolution.

For a given ``binding'' potential $V(q)$ the angular frequency $\omega$ is generally defined as one half of the 2nd derivative of 
$V(q)$ at its (local) minimum $q=q_0$ : $\omega = (1/2)\,d^2V(q)/dq^2_{|q=q_0}$.
\section{Quantum mechanics of the phase space $\mathcal{S}_{\vp,I}$}
\subsection{Basics: self-adjoint representations of the \\ three Lie algebra  generators $K_j$  of \\ the symplectic group $Sp(2,\mathbb{R})$ }
The quantization of the global ``coordinates'' $h_j$ from Eq.\
\eqref{eq:11} is implemented by reinterpreting them as self-adjoint
operators in a given Hilbert space \cite{ka12},  \begin{equation}
 h_j \to K_j=\hbar\,\tl{K}_j\, \label{eq:104}
 \end{equation} which obey the associated Lie algebra
 \eqref{eq:12}:
 \begin{align} \label{eq:105}
  [\tl{K}_0,\,\tl{K}_1] = i\,\tl{K}_2\,,&~~[\tl{K}_0,\,\tl{K}_2] =
  -i\,\tl{K}_1\,, \\
  [\tl{K}_1,\,\tl{K}_2]& = -i\,\tl{K}_0\,.
\end{align}
(Quantities $\tl{A}$ with a "tilde", here and in the following, are considered to be dimensionless).

The three self-adjoint operators $\tl{K}_j$ can be obtained as Lie algbra generators of
irreducible  unitary  representations of the corre\-sponding groups
  $SO^{\uparrow}(1,2)$, $ Sp(2, \mathbb{R})$ (the latter being isomorphic to the groups  $ SL(2, \mathbb{R})$ and $
 SU(1,1)$) or of one of their infinitely many covering groups \cite{kas3}.

As $\tl{K}_0$ is the generator of the maximal compact abelian
subgroup $U(1)$, its eigenstates may be used as a Hilbert space basis (here formally in Dirac's notation, explicit examples will be discussed later):
\[
\tl{K}_0 |b,\,n \rangle = (n+b)\,|b,\,n \rangle\,, \] where $b$
is some real number $\in (0,1)$  (``Bargmann index'' \cite{barg}) and $n= 0,1,2,\ldots$. This central result can be derived as follows

 The operators \begin{align} ~\tl{K}_{\pm}
=& \tl{K}_1 \label{eq:34}
   \pm i\,\tl{K}_2\,, \\ &K_1 =\frac{1}{2}(\tl{K}_+ + \tl{K}_-),\,K_2 =\frac{1}{2i}(\tl{K}_+ -\tl{K}_-),\label{eq:201}
 \end{align}
 obey the relations
\begin{align} \label{eq:60}
  [\tl{K}_0,\,\tl{K}_+] = \tl{K}_+\,,&~~[\tl{K}_0,\,\tl{K}_-] =
  -\tl{K}_-\,, \\
  [\tl{K}_+,\,\tl{K}_-]& = -2\tl{K}_0\,. \label{eq:229}
\end{align}
They are raising and lowering operators:  \begin{align}
\label{eq:35}
  \tl{K}_+ |b,\,n \rangle& = [(2b+n)(n+1)]^{1/2}\,|b,\,n+1 \rangle\,,\\ \tl{K}_- |b,\,n \rangle& =
 [(2b+n-1)n]^{1/2}\,|b,\,n-1 \rangle\,.\label{eq:52}
\end{align} 

The relations \eqref{eq:35} and \eqref{eq:52} are derived under the assumptions that there exists a state $|b,0 \rangle$ such that \begin{equation}
\tl{K}_0|b,0\rangle = b\,|b,0 \rangle\,,~~~\tl{K}_- |b,0\rangle =
0,~ b \in [1,0). \end{equation}
Eq.\ \eqref{eq:35} implies
\begin{align}
  \label{eq:237}
  |b,n \rangle  = & \frac{1}{\sqrt{(2b)_n\,n!}}(\tl{K}_+)^n|b,0\rangle,\\ &(2b)_n = 2b(2b+1)\cdots (2b + n -1) \nonumber \\ &~~~~~~~ = \Gamma(2b+n)/\Gamma(2b). \nonumber
\end{align}

It then follows that
\begin{equation}\label{eq:36}
  \tl{K}_0\,|b,n\rangle =
  (n+b)\,|b,n\rangle\,,\,n=0,1,\ldots;~~1>b>0\,.
\end{equation}This is the so-called ``positive discrete series'' $D_b^{(+)}$
among the different types of possible irreducible unitary
representations of $Sp(2,\mathbb{R})$ \cite{barg,kas3}  The Bargmann index  $b$ - called ``B-index'' in the following -  characterizes an irreducible unitary representation (IUR)
$D_b^{(+)}$.

The Casimir operator
\begin{align}\label{eq:65}
  \mathfrak{C}&=\tl{K}_1^2+\tl{K}_2^2-\tl{K}_0^2 \\ \nonumber
  &= \tl{K}_+\tl{K}_- - \tl{K}_0\,(\tl{K}_0 -1) \\ \nonumber
  &= \tl{K}_-\tl{K}_+ - \tl{K}_0\,(\tl{K}_0+ 1)
\end{align}
of the IUR $D_b^{(+)}$ has the value \begin{equation}
  \mathfrak{C} = b(1-b){\bf 1}\,.\label{eq:42} \end{equation}

This means that the ``classical Pythagoras'' \eqref{eq:228} is
violated  quantum mechanically for $b\neq 1$, e.g.\ in the case of the usual HO
with $b=1/2$!

The Group $SO^{\uparrow}(1,2)$ has {\em infinitely many covering
groups} because its compact subgroup   $O(2)\cong S^1 \cong U(1)$
 is infinitely connected!

 Let us denote the
 $s$-fold covering by
\begin{equation} \label{eq:37} SO^{\uparrow}_{[s]}(1,2)\,,\,s=1,2,\ldots\,.\end{equation} Its
irreducible unitary representations $D_b^{(+)}$ have the indices
\begin{equation}\label{eq:38}
  b= \frac{r}{s}\,,~r =1,2,\ldots, s-1\,.
\end{equation}
 This means that $ b_{min}=1/s $ can be arbitrarily small $>0$ if $s$ is large enough!  \\
The 2-fold coverings \begin{equation} Sp(2,\mathbb{R})\label{eq:39}
  =SL(2,\mathbb{R})\,\cong SU(1,1) \end{equation} have $ b= 1/2$.

 The results above imply that the  $(\vp,I)$-Hamiltonian
\begin{equation} \label{eq.:40}
  H(\vp,I) \to H_{osc}(K) =  \omega\,K_0\,,~~K = \hbar\,(\tl{K}_0,\tl{K}_1,\tl{K}_2)\,
\end{equation}
can have the $b$-dependent spectra  \begin{equation}
  E_{b,\,n}(\vp,I) = \hbar\,\omega\, (n+b)\,,\, n=0,1,\ldots\,;~~1>b>0, \label{eq:41}
\end{equation}
As this result is due to the properties of the symplectic group $Sp(2, \mathbb{R})$, especially its compact subgroup $U(1) \cong O(2)$, we call it
the ``symplectic spectrum'' of the HO, and the conventional special case $b=1/2$ as its ``orthodox'' one.

Because of the relations \eqref{eq:201}, \eqref{eq:35} and \eqref{eq:52} one has
\begin{equation}
  \langle b,n|\tl{K}_j|b,n \rangle = 0\,,~j=1,2,\label{eq:61}
\end{equation}
implying for the mean square  deviations
\begin{equation}\label{eq:62}
  (\Delta \tl{K}_j)^2_{b,n} =
\frac{1}{2}(n^2+2nb+b)\,,~j=1,2,
\end{equation} so that
\begin{equation} \label{eq:63}
 (\Delta \tl{K}_1)_{b,n}\,(\Delta \tl{K}_2)_{b,n} = \frac{1}{2}(n^2+2b n+b)\,, \end{equation}  \begin{equation} \label{eq:64} (\Delta \tl{K}_1)_{b,n=0}\,
(\Delta \tl{K}_2)_{b,n=0} =
 \frac{b}{2}\,.
\end{equation} The last relation shows that $b\to 0$ is a kind of  classical
limit in the angle-action framework!

\subsection{Time evolution}
\subsubsection{Heisenberg picture}
The appearence of covering groups \eqref{eq:37} has a natural physical background:
Take the time dependence of the angle $\vp(t) = \omega\,t$ in Eq.\ \eqref{eq:6} (with $t_0=0,\vp_0=0$):
In general the system will not stop after covering a circle just once, $\omega\,T_1 = 2\pi$, but will
circle the origin, say, at least $s$ times, $\omega\,T_s = 2\pi\,s$, where $s \in \mathbb{N}$ can be arbitrarily large.

In the following the dimensionless time variable 
\begin{equation}
  \label{eq:68}
  \tl{t} = \omega\,t
\end{equation}
will be used. It is an angle variable.

 The unitary time evolution operator is given by
\begin{equation} \label{eq:43}
  U(\tl{t}) = e^{-i\,\tl{K_0}\,\tl{t}}\,,~~  \tl{K}_0=N+b\bf{1}\,,
\end{equation}
where the number operator $N$ can be considered a function of the operators $\tl{K}_j,\, j=0,1,2$, as will be shown below.

The unitary operator \eqref{eq:43} implies the usual Heisenberg
Eqs.\ of motion:
\begin{align}
 U(-\tl{t})\,\tl{K}_+\,U(\tl{t}) &= e^{-i\tl{t}}\tl{K}_+,  
                                   \label{eq:44}\\
   U(-\tl{t})\,\tl{K}_-\,U(\tl{t}) &= e^{+i\tl{t}}\tl{K}_-,  \label{eq:202} \\
 U(-\tl{t})\,\tl{K}_1\,U(\tl{t}) &= \cos\tl{t}\,\tl{K}_1 +
 \sin\tl{t}\,\tl{K}_2, \label{eq:203} \\
 U(-\tl{t})\,\tl{K}_2\,U(\tl{t}) &=-\sin\tl{t}\,\tl{K}_1 +
 \cos\tl{t}\,\tl{K}_2.\label{eq:67}
\end{align}
 For $\tl{t} = 2\pi$ the operator
\eqref{eq:43} becomes
\begin{equation}\label{eq:45}
  U(\tl{t} = 2\pi) = e^{-2\pi i b}{\bf 1}\,.
\end{equation} If
\begin{equation} b= r/s\,,~r,s \in \mathbb{N},\, r<s \text{ and divisor free}, \label{eq:46} \end{equation} this implies for
$SO^{\uparrow}_{[s]}(1,2)$:
\begin{equation}\label{eq:47} U(\tl{t} = s\,2\pi)= {\bf
    1}\,.\end{equation}
  The ground state $|b,0\rangle$ has the time evolution
\begin{equation}\label{eq:48}
  U(\tl{t})\,|b,0\rangle = e^{-i\,b\,\tl{t}}\,|b,0 \rangle \,,
\end{equation} with the associated
time period  \begin{equation}\label{eq:49}
  T_{2\pi, b} = \frac{2\pi}{\omega_b},\;\; \omega_b\equiv  b\,\omega\,,
\end{equation} which
can become arbitrarily large for $b=1/s,\, s\to \infty$.
Symbolically speaking: $1/b$ is a kind of refraction index $n$.

Whereas $\tl{K}_0$ generates rotations and time evolutions by performing many phase rotations,
the operators $\tl{K}_1$ and $\tl{K}_2$ generate special ``Lorentz'' transformations (``boosts'') in directions 1 and 2, respectively \cite{ka16}:

With
\begin{align}
  \label{eq:152}
  U(w) =& e^{(w/2)\tl{K}_+ - (w^{\ast}/2)\tl{K}_-} = e^{i\,w_2\tl{K}_1 + i\,w_1\tl{K}_2}, \\
  & w= w_1 + i\,w_2 =|w|e^{i\theta} \nonumber
\end{align}
one obtains
\begin{align}\label{eq:153}
  U(-w)\vec{\tl{K}}U(w) =& \vec{\tl{K}} (\cosh|w| -1)(\vec{n}\cdot\vec{\tl{K}})\cdot \vec{n} \\ & + \sinh|w|\,\vec{n}\,\tl{K}_0, \nonumber \\
  U(-w)\tl{K}_0U(w) = & \cosh|w|\tl{K}_0 + \sinh|w| (\vec{n}\cdot\vec{\tl{K}}), \label{eq:154} \\
  & \vec{\tl{K}} = (\tl{K}_1,\tl{K}_2),\; \vec{n} = (\cos\theta, -\sin\theta). \nonumber
\end{align}
Eqs.\ \eqref{eq:153} and \eqref{eq:154} describe a Lorentz ``boost'' in direction $\vec{n}$.
\subsubsection{Schr\"odinger Picture}
If $|b;t_0\rangle $ is a state vector of the Hilbert space associated with the B-index $b$ at time $t_0$, then - according to the unitary time evolution
\eqref{eq:43} - we have at time $t$:
\begin{equation}
  \label{eq:204}
  |b;t\rangle = U(t-t_0)|b;t_0\rangle =e^{-i\omega K_0 (t-t_0)/\hbar}|b;t_0\rangle,
\end{equation}
which implies
\begin{equation}
  \label{eq:205}
  \frac{\hbar}{i}\partial_t|b;t\rangle =\omega K_0|b;t\rangle,~K_0 =\hbar \tl{K}_0.  
\end{equation} 
For a representation
\begin{equation}
  \label{eq:206}
  |b;t\rangle = \sum_{n=0}^{\infty}c_n(t)|b,n\rangle
\end{equation}
we get from Eq.\ \eqref{eq:204}, with $t_0=0$,
\begin{equation}
  \label{eq:207}
  c_n(t) = e^{-i\omega(n+b)t},
\end{equation}
so that
\begin{equation}
  \label{eq:208}
|b;t\rangle = e^{-ib\, \omega t}\sum_{n=0}^{\infty}e^{-i\,n\,\omega t}|b,n\rangle.
\end{equation}
\subsection{Relationship between the operators $\tl{K}_j$ \newline and the conventional operators $Q$ and $P$}
The relations \eqref{eq:1} expressed in terms of the functions $h_j(\vp,I)$ from Eqs.\ \eqref{eq:11} take the form
\begin{align}\label{eq:50}
  q(\vp,I)&=
  \sqrt{\frac{2}{m\,\omega}}\,\frac{h_1(\vp,I)}{\sqrt{h_0(\vp,I)}}\,, \\
 p(\vp,I)&= \sqrt{2\,m\,\omega}\,\frac{h_2(\vp,I)}{\sqrt{h_0(\vp,I)}}\,.
\end{align}
There is a corresponding relationship at the operator level: Define  the operators
 \begin{align} \label{eq:51}
  A(K)  &= (\tl{K}_0+b)^{-1/2}\tl{K}_-, \\  A^{\dagger}(K)& =  \tl{K}_+(\tl{K}_0+b)^{-1/2}. \label{eq:231}
\end{align}
According to Eqs.\ \eqref{eq:36}, \eqref{eq:35} and \eqref{eq:52} they act on the number states $|b,n\rangle$ as
\begin{align}\label{eq:53}
  A^{\dagger}\,|b,n \rangle& =  \sqrt{n+1}\,|b, n+1 \rangle, \\
 A\,|b,\,n \rangle & =  \sqrt{n}\,|b,n-1 \rangle,\label{eq:230} \\ &\, n= 0, 1, \ldots\,. \nonumber
\end{align}
This means \begin{equation}\label{eq:54}
 [A,\,A^{\dagger}] = {\bf 1}\, ~~  \forall~ D_b^{(+)},
\end{equation}
{\em independent of the value of $b$~!}

Thus, the composite operators $A(K)$ and $A^{\dagger}(K)$ are the usual Fock space annihilation and creation operators for all $b$ and independent of $b$~!

Note that the denominator in Eqs.\ \eqref{eq:51} and \eqref{eq:231} is well-defined, because $\tl{K}_0$ is a positive definite operator and $b$ a
positive number for each representation of the series $D^{(+)}_b$.

The quantum mechanical position and momentum operators $Q$ and $P$ can now be defined as usual:
\begin{align} \label{eq:55}
  Q=Q(K)=&\, \frac{\lambda_0}{\sqrt{2}}\,[A^{\dagger}(K)+A(K)], \\ P=P(K)=&\,
 \frac{i\,\hbar}{\sqrt{2}\,\lambda_0}\,[A^{\dagger}(K)-A(K],\label{eq:56} \\ &\, \lambda_0 =
\sqrt{\hbar/(m\,\omega)},\label{eq:57} 
\end{align}
where $\lambda_0$ has the dimension of a length.

The $(q,p)$-Hamilton operator
\begin{align}\label{eq:58}
  H(Q,P) =\, & \frac{1}{2m}P^2(K) + \frac{1}{2}m\,\omega^2Q^2(K)\\ =\,& \hbar\,\omega(N(K) + \frac{1}{2} {\bf 1}),\nonumber \\ &\, N(K) = A^{\dagger}(K)A(K) \label{eq:59}
\\ &~~~~~~~~ = \tl{K}_+(\tl{K}_0+b)^{-1}\tl{K}_-  \nonumber \end{align}
has the usual ``orthodox'' spectrum $E_n = \hbar \omega (n + 1/2)$\:!

Thus, it turns out that the quantum mechanics associated with the phase space $\mathcal{S}_{\vp,I}$ is rather more subtle than the one associated with
$\mathcal{S}_{q,p}$ and that those subtleties get lost if one passes from the $(\vp,I)$-case to the $(q,p)$-case!

As the creation and annihilation operators
$A^{\dagger}$ $A$ with their ($b$-independent) defining properties \eqref{eq:53}, \eqref{eq:230} and \eqref{eq:54} are essential building blocks for many quantum systems, that loss
of $b$-dependent subtleties may in turn lead to a corresponding loss of physical insights. \emph{ The big question is: Did nature implement those subtleties?}

The time evolution of the composite operators \eqref{eq:55} and \eqref{eq:56} is the usual one. It follows from the relations \eqref{eq:44} and \eqref{eq:67}:
\begin{align}
U(-\tl{t})\,\tl{Q}\,U(\tl{t}) =& \cos\tl{t}\,\tl{Q} + \sin\tl{t}\,\tl{P}\,,\label{eq:66} \\
 U(-\tl{t})\,\tl{P}\,U(\tl{t}) =& -\sin\tl{t}\,\tl{Q} +
\cos\tl{t}\,\tl{P}\,.
\end{align}
\subsection{The model $H=\hbar\,\omega(\tl{K}_0 +g\,\tl{K}_1)$}
\subsubsection{Transition matrix elements with respect to  the number states in 1st order}
The quantum mechanical counterpart of the classical Hamiltonian \eqref{eq:14} is the operator
\begin{equation}
  \label{eq:69}
  H(K) = \hbar\,\omega\,\tl{C}_g(K),\, \tl{C}_g(K) = \tl{K}_0 + g\,\tl{K}_1,\,g \ge 0\,.
\end{equation}

Before discussing a special explicit choice for the operators $\tl{K}_j$, their associated Hilbert space and the exact eigenfunctions of the Hamiltonian \eqref{eq:69} we mention the values of the (formal) 1st order matrix
elements
\begin{equation}
  \label{eq:70}
  \langle b,m|\tl{C}_g(K)|b,n\rangle,\;m,n = 0,1,2, \ldots
\end{equation}

From the relations \eqref{eq:201}, \eqref{eq:35} and \eqref{eq:52} we get for $m \neq n$ (the case $m=n$ appears trivial, but that is  only so in 1st order. It follows from the exact solution - discussed below - that  the 2nd order $g^2$ and higher ones
contribute):
\begin{align}\label{eq:72}
  \langle b,m|\tl{C}_g(K)|b,n\rangle =& \frac{g}{2}\langle b,m|(\tl{K}_+ + \tl{K}_-)|b,n\rangle \\
  =& \frac{g}{2} [(2b+n)(n+1)]^{1/2}\,\delta_{m (n+1)} \nonumber \\
  +& \frac{g}{2} [(2b+n-1)n]^{1/2}\,\delta_{m (n-1)}. \nonumber
\end{align}
Thus, we have the selection rule
\begin{equation}
  \label{eq:114}
  \Delta n = \pm 1,
\end{equation}
for the Hamiltonian \eqref{eq:69}, 
the same as, e.g., for vibrational (electric dipole) transitions of diatomic molecules \cite{herz1}!

Examples:
\begin{align}\label{eq:73}
  \langle b, m=0|\tl{C}_g(K)|b, n=1\rangle =& \frac{g}{2} \sqrt{2b}, \\
  \langle b, m=2|\tl{C}_g(K)|b. n=1\rangle =& \frac{g}{2}\sqrt{2(2b+1)}, \label{eq:74} \\ 
  \langle b, m=n_0-1 |\tl{C}_g(K)|b, n_0\rangle =& \frac{g}{2}\sqrt{n_0(2b+n_0-1)}, \label{eq:75} \\
  \langle b, m=n_0+1|\tl{C}_g(K)|b, n_0\rangle =& \frac{g}{2}\sqrt{(n_0+1)(2b+n_0)} \label{eq:76}
\end{align}
Eq.\ \eqref{eq:73} shows that the associated transition probability for $0 \leftrightarrow 1$ is proportional to $b$.

\subsubsection{Exact eigenvalues of $\tl{C}_g(K)$}
It follows from the explicit Hilbert space calculations in Ch.\ III.E and in Appendix A that the operator $\tl{C}_g(K)$ has the exact eigenvalues
\begin{equation}
  \label{eq:33}
  \tl{c}_{g,b;n} = (n+b)\sqrt{1-g^2},~n=0,1,\ldots,
\end{equation}
so that the Hamiltonian \eqref{eq:69} has the the eigenvalues
\begin{equation}
  \label{eq:283}
  \tl{E}_{g,b;n} = \hbar \omega_g(n+b),~n=0,1,\ldots,~\omega_g=\sqrt{1-g^2}\omega,
\end{equation}
which reflects the ("redshifted") frequency reduction \eqref{eq:93} of  the classical motions in Ch.\ II.
\subsection{Explicit Hilbert spaces for $\tl{K}_j, j= 0,1,2$ and $ \tl{C}_g(K) = \tl{K}_0 + g\tl{K}_1$,  spectra  and eigenfunctions}
Several explicit Hilbert spaces for concrete irreducible unitary representations of the group $SO^{\uparrow}(1,2)$, its twofold covering the symplectic group $Sp(2,\mathbb{R})$ (or the isomorphic ones  $SU(1,1)$ and
$SL(2,\mathbb{R}$) and of all other covering groups as well  have been
discussed in the literature \cite{barg,sall1,vile,sall2,wolf,ka13}.

The associated self-adjoint Lie algebra generators $\tl{K}_j$ all obey the same commutation relations \eqref{eq:32}. The representation spaces include Hardy spaces on the unit circle $\mathbb{S}^1$, Hilbert spaces of holomorphic functions on the unit disc
$ \mathbb{D} = \{\lambda \in \mathbb{C}, |\lambda| < 1\}$ and also Hilbert spaces on the positive real line $\mathbb{R}^+_0 = \{x \in \mathbb{R}, x \ge 0\}$.
We shall present Hardy space related Hilbert spaces for $b \in (0,1)$ here and discuss corresponding Hilbert spaces on $\mathbb{R}^+_0$ in  Appendix A:
Hardy spaces on the unit circle are closely related to the variable angle, whereas Hardy Hilbert spaces on $\mathbb{R}^+_0$ are associated with the action variable $I$.

The following discussion follows closely those of Secs.\ 7.1 and 7.2 of Ref.\ \cite{ka8}. Mathematical details like,  e.g.\ questions concerning the convergence of series or integrals, will be ignored in the following! The associated justification can be found in the mathematical literature quoted above.
\subsubsection{Hardy space on the unit circle}
A ``Hardy space'' $H^2_+(\mathbb{S}^1, d\vt)$ is a closed subspace of the usual Hilbert space $L^2(\mathbb{S}^1, d\vt)$ on the
unit circle $\mathbb{S}^1$ with the scalar product
\begin{equation}
 \label{eq:155}
  (f_2,f_1) = \frac{1}{2\pi}\,\int_{S^1}d\vt\,f_2^*(\vt)f_1(\vt)\,,
\end{equation}
and the orthonormal basis
\begin{equation}
  \label{eq:156}
 e_n(\vt) = e^{i\,n\,\vt}\,,~~ n \in \mathbb{Z}\,.
\end{equation}
The associated Hardy subspace $H^2_+(\mathbb{S}^1,\,d\vt)$ is spanned by the basis consisting of the elements with non-negative $n$,
namely
\begin{equation}
  \label{eq:157}
  e_n(\vt)= e^{i\,n\,\vt}\,,~~n=0,1,2,\cdots\,.
\end{equation}

 If we have  two Fourier series $\in H^2_+(\mathbb{S}^1\,,d\vt)$,
\begin{equation}
  \label{eq:158}
 f_1(\vt)=\sum_{n=0}^{\infty} c_{n,1}\, e^{i\,n\,\vt}\,,~
~f_2(\vt)= \sum_{n=0}^{\infty}c_{n,2}\,e^{ i\,n\,\vt}\,,
\end{equation}
they have the scalar product
\begin{equation}
 \label{eq:159}
 (f_2,f_1)_+=
\frac{1}{2\pi}\int_{S^1} d\vt\,
f^*_2(\vt)f_1(\vt)=\sum_{n=0}^{\infty}c^*_{n,2}\,c_{n,1}
\end{equation}
and obey the boundary condition
\begin{equation}
  \label{eq:210}
  f_j(\vt + 2\pi) = f_j(\vt), \; j=1,2.
\end{equation}
The coefficients $c_{n,j}$ are given by
\begin{equation}
  \label{eq:257}
  c_{n.j} = (e_n,f_j)_+.
\end{equation}
 $Sp(2, \mathbb{R})$ Lie algebra generators  are 
\begin{eqnarray}
\tl{K}_0 &=& \frac{1}{i}\partial_{\vt} + \frac{1}{2}\,,\label{eq:160} \\
\tl{K}_+ &=&
e^{i\,\vt}\,(\frac{1}{i}\partial_{\vt}+1), \label{eq:161} \\
\tl{K}_- & =& e^{-i\,\vt}\,\frac{1}{i}\partial_{\vt}. \label{eq:162}
\end{eqnarray}
Thus, the Hardy space with the basis \eqref{eq:157} provides a Hilbert space for the conventional HO with the spectrum $\{n+ 1/2$\} and the operators
\eqref{eq:160}-\eqref{eq:162} act on the basis \eqref{eq:157} as
\begin{eqnarray}
\tl{K}_0\,e_n(\vt) &=& (n+\frac{1}{2})\,e_n(\vt)\,, \label{eq:163}\\
\tl{K}_+\,e_n(\vt)&=& (n+1)\,e_{n+1}(\vt)\,, \label{eq:164} \\
\tl{K}_-\,e_n(\vt)&=& n\,e_{n-1}(\vt)\,, \label{eq:165}\end{eqnarray}
which are special cases of the relations \eqref{eq:36}, \eqref{eq:35} and \eqref{eq:52} with $b=1/2$.
Note that the ground state here is given by $e_{n=0}(\vt) = 1$.

A possible generalization of the case $b=1/2$ for $b \in (0,1)$ within the same Hilbert space as above can be obtained \cite{boj,hol} by inspection of the relations \eqref{eq:36}, \eqref{eq:35} and \eqref{eq:52}:
\begin{align}
  \tl{K}_0 &= N + b,~~N= \frac{1}{i}\partial_{\vt}, \label{eq:258} \\
  \tl{K}_+ & = e^{i\vt}[(N+2b)(N+1)]^{1/2}, \label{eq:259} \\
   \tl{K}_- & = [(N+2b)(N+1)]^{1/2}e^{-i\vt}.\label{eq:260}
\end{align}
Applied to the basis \eqref{eq:157} these operators have the correct properties.

Defining the  self-adjoint operator
  \begin{align}
    M_b(N)& =  +[(N+2b)(N+1)]^{1/2}  =M_b^{\dagger},         \label{eq:262}   \\
    & = +[(\tl{K}_o+b)(N+1)]^{1/2} , \nonumber
  \end{align} with
  \begin{equation}
    \label{eq:264}
    M_b(N)\,e^{i n \vt} = +[(n+2b)(n+1)]^{1/2}\,e^{i n \vt},
  \end{equation}
  the operators \eqref{eq:259} and \eqref{eq:260} can be written as
  \begin{equation}
    \label{eq:263}
    \tl{K}_+ = (e^{i\vt}\,M_b)\,,~~ \tl{K}_- = (M_b\,e^{-i\vt}).
  \end{equation}

\subsubsection{Hardy space related Hilbert spaces for general $b \in (0,1]$}
Another possible representation for the  more general case $b \in (0,1) $ can be obtained by a generalization of the the scalar product \eqref{eq:159}:

Introducing on $H_+^2(\mathbb{S}^1,d\vt)$
 the positive definite (self-adjoint) operator $A_b$ by the  action
\begin{align}
  \label{eq:166}
  A_b\,e_n(\vt)=& \frac{n!}{(2b)_n}e_n(\vt),~n=0,1,\ldots\,,~b>0, \\
  & (a)_n = a\,(a+1)\,(a+2)\ldots (a+n-1)\nonumber \\ &~~~~~\; = \Gamma(a+n)/\Gamma(a),\,\,(a)_{n=0}=1, \nonumber
\end{align}
 one can define an additional scalar product for functions
\begin{equation}
  \label{eq:167}
  f_j(\vt)=\sum_{n=0}^{\infty}c_{n,j}e_n(\vt)\,,~j=1,2,
\end{equation}
by
\begin{equation}
  \label{eq:168}
  (f_2,f_1)_{b,+} \equiv (f_2,A_b\,f_1)_+ = \sum_{n=0}^{\infty}\frac{n!}{(2b)_n}\,c^*_{n,2}\,c_{n,1}\,,
\end{equation}
so that
\begin{equation}
  \label{eq:169}
 (e_n,f_1)_{b,+} =\frac{n!}{(2b)_n}\,c_{n,1}. 
\end{equation}
We denote the (Hardy space associated) Hilbert space with the scalar product \eqref{eq:168} by $H^2_{b,+}(\mathbb{S}^1,d\vt)$.

An orthonormal basis in this space is given by
\begin{align}
  \label{eq:170}
  \hat{e}_{b,n}(\vt)=& \sqrt{\frac{(2b)_n}{n!}}\,e_n(\vt)\,, \\ &( \hat{e}_{b,n_2}, \hat{e}_{b,n_1})_{b,+} = \delta_{n_2n_1}.\nonumber
\end{align}  Two series
\begin{equation}
  \label{eq:171}
  f_j(\vt)= \sum_{n=0}^{\infty}a_{n,j}\,\hat{e}_{b,n}(\vt)\,,~j=1,2,
\end{equation}
have the obvious scalar product
\begin{align}
  (f_2,f_1)_{b,+}=&\sum_{n=0}^{\infty}a^*_{n,2}\,a_{n,1}, \label{eq:172} \\
a_{n,1} =  & (\hat{e}_{b,n},f_1)_{b,+}  \label{eq:173}
\end{align}
It follows that for a given function $f(\vt)$ its expansion coefficients $c_n$ or $a_n$ with respect to $e_n$ or $e_{b,n}$ are related as follows
\begin{equation}
  \label{eq:261}
  c_n = (e_n,f)_+ = \sqrt{\frac{n!}{(2b)_n}}\,(\hat{e}_{b,n},f)_+ =\sqrt{\frac{n!}{(2b)_n}}\,a_n
\end{equation}
As in general
\begin{equation}
  \label{eq:265}
  (f,f)_+ \ne (f,f)_{b,+},
\end{equation}
one has to be careful in the case of quantum mechanical applications:

In a Hilbert space with scalar product $(f_2,f_1)$ one generally needs the normalization $(f,f)=1$ for the usual probability interpretations. If initially $(f,f) \ne 1$
one has to renormalize the state $f$: $f \to f/\sqrt{(f,f)}$. So, if, e.g. $(f,f)_+ = 1$ in  inequality \eqref{eq:265}, one has to renormalize $f$ if one wants to
determine transition probabilities and expectation values etc. with respect to \eqref{eq:168}.

In $H^2_{b,+}(\mathbb{S}^1,d\vt)$ the generators $\tl{K}_j$ have the form \cite{ka12,ka13}
\begin{align}
  \label{eq:174}
  \tl{K}_0 =& \frac{1}{i}\,\partial_{\vt}+ b, \\ \tl{K}_+=& e^{i\,\vt}(\frac{1}{i}\partial_{\vt}+ 2b), \label{eq:191} \\ \tl{K}_-=& e^{-i\,\vt}
\frac{1}{i}\partial_{\vt}, \label{eq:192}
\end{align}
so that
\begin{align}
  \tl{K}_1 =& \frac{1}{2}(\tl{K}_+ +\tl{K}_-) = \cos\vt\,\frac{1}{i}\partial_{\vt} + b\,e^{i\vt}, \label{eq:193} \\
  \tl{K}_2 =& \frac{1}{2i}(\tl{K}_+ -\tl{K}_-) = \sin\vt\,\frac{1}{i}\partial_{\vt} -ib\,e^{i\vt} \label{eq:194} \\ =& \tl{K}_1(\vt-\pi /2). \nonumber
\end{align}
The operators \eqref{eq:191} and \eqref{eq:192} have the correct actions \eqref{eq:35} and \eqref{eq:52} on the basis \eqref{eq:170}:
\begin{eqnarray}
  \label{eq:175}
  \tl{K}_0\,\hat{e}_{b,n}&=& (n+b)\,\hat{e}_{b,n}\,,\\ \tl{K}_+\,\hat{e}_{b,n}&=& \sqrt{(2b+n)(n+1)}\,\hat{e}_{b,n+1}\,,
\label{eq:176}
\\ \tl{K}_-\,\hat{e}_{b,n} &=& \sqrt{(2b+n-1)n}\,\hat{e}_{b,n-1}\,. \label{eq:177}
\end{eqnarray}
The operators \eqref{eq:191} and \eqref{eq:192} 
are adjoint to each other only with respect to the scalar product \eqref{eq:168}, not with respect to \eqref{eq:159}.
Their adjointness with respect  to the scalar product \eqref{eq:168} can be verified by taking two series \eqref{eq:171}
and showing that
\begin{equation}
  \label{eq:256}
  (\tl{K}_-f_2,f_1)_{b,+} = (f_2,\tl{K}_+f_1)_{b,+}.
\end{equation}
This relation implies the self-adjointness of the operators \eqref{eq:193} and \eqref{eq:194}.
Note that
\begin{align}
  \label{eq:178}
  (e_{n_2},\hat{e}_{b,n_1})_+  &= (\hat{e}_{b,n_1},e_{n_2})_+ = \sqrt{\frac{(2b)_{n_1}}{n_1!}}\,\delta_{n_2\,n_1}, \\
(e_{n_2},\hat{e}_{b,n_1})_{b,+}  &= (\hat{e}_{b,n_1},e_{n_2})_{b,+} = \sqrt{\frac{n_1!}{(2b)_{n_1}}}\,\delta_{n_2\,n_1},
\label{eq:179} \\(\hat{e}_{b,n_2},\hat{e}_{b,n_1})_+  &= \frac{(2b)_{n_1}}{n_1!}\,\delta_{n_2\,n_1}, \label{eq:190}  \\
 (e_{n_1},e_{n_2})_{b,+}  &= \frac{n_1!}{(2b)_{n_1}}\,\delta_{n_2\,n_1}\,.\label{eq:180}
\end{align}

The Fock space ladder operators $A$ and $A^{\dagger}$ associated with the Lie algebra generators \eqref{eq:174}-\eqref{eq:192} are
given according to Eqs.\ \eqref{eq:51}. 

The so-called ``reproducing kernel'' on $H^2_{b,+}$ is given by the ``completeness'' relation
\begin{align}
  \label{eq:181}
  \hat{A}_b(\vt_2-\vt_1) = & \sum_{n=0}^{\infty}\hat{e}^*_{b,n}(\vt_2)\,\hat{e}_{b,n}(\vt_1) \\ =
 [1-e^{ i\,(\vt_1-\vt_2)}]^{-2b}= & \hat{A}_b^*(\vt_1-\vt_2)\,,
\end{align}
where the identity
\begin{equation}
  \label{eq:211}
\frac{(a)_n}{n!}  = (-1)^n {- a \choose n}
\end{equation}
has been used.
According to the relations \eqref{eq:178} -- \eqref{eq:180} the kernel \eqref{eq:181} has the properties
\begin{equation}
  \label{eq:212}
 \frac{1}{2\pi} \int_{\mathbb{S}^1}d\vt_2 \hat{A}_b(\vt_2-\vt_1)\hat{e}_{b,m}(\vt_2) = \hat{e}_{b,m}(\vt_1),
\end{equation}
or, written more formally in terms of the scalar product \eqref{eq:168}
\begin{align}  
  (\hat{A}_b(1,2),\,\hat{e}_{b,m}(2))_{b,+} &= \hat{e}_{b,m}(\vt_1), \label{eq:182} \\
 (\hat{A}_b(1,2),\,\hat{e}_{b,m}(2))_{+} &= \frac{(2b)_m}{m!}\, \hat{e}_{b,m}(\vt_1), \label{eq:183} \\
(\hat{A}_b(1,2),\,e_{m}(2))_{b,+} &= \sqrt{\frac{m!}{(2b)_m}}\,\hat{e}_{b,m}(\vt_1)= e_m(\vt_1),\label{eq:184} \\
(\hat{A}_b(1,2),\,e_{m}(2))_{+} &= \sqrt{\frac{(2b)_m}{m!}}\,\hat{e}_{b,m}(\vt_1)\label{eq:185} \\   &= \frac{(2b)_m}{m!}\,e_m(\vt_1).
\nonumber
\end{align}
The numbers $1$ and $2$ mean the variables $\vt_1$ and $\vt_2$, the latter being an integration variable.

The scalar product \eqref{eq:168} itself may - according to Eq.\ \eqref{eq:185} - be written as
\begin{equation}
  \label{eq:213}
 \frac{1}{(2\pi)^2} \int_{\mathbb{S}^1}d\vt_2 \int_{\mathbb{S}^1}d\vt_1 f_2^{\ast}(\vt_2)A_b(\vt_2-\vt_1)f_1(\vt_1),
\end{equation}
where the functions $f_j(\vt)$ are as in Eq.\ \eqref{eq:171}.
If a function
\begin{equation}
  \label{eq:71}
  f(\vt_2) = \sum_{n=0}^{\infty} a_n\hat{e}_{b,n}(\vt_2)
\end{equation}
is an element of $H^2_{b,+}$ then it follows from \eqref{eq:182} that
\begin{equation}
  \label{eq:209}
  (\hat{A}_b(1,2),\,f(2))_{b,+} = f(\vt_1).
\end{equation}
Thus, the ``reproducing kernel'' $\hat{A}_b(2,1)$ has - formally - similar properties as the usual $\delta$-function.

The property \eqref{eq:209} has the following calculational advantage: If one has two functions \eqref{eq:171} considered as elements of  $H^2_{b,+}$, then their
scalar product \eqref{eq:168} can be calculated as
\begin{equation}
  \label{eq:270}
  (f_2,f_1)_{b,+} = \frac{1}{2\pi}\int_{\mathbb{S}^1}\, d\vt \,f_2^{\ast}(\vt)f_1(\vt).
\end{equation}

{\bf Space reflection and time reversal}

According to Eq.\ \eqref{eq:1} space reflections $P$ can be implemented by the substitution
\begin{equation}
  \label{eq:218}
  P:~~ \vt \to \vt + \pi
\end{equation}
and time reversal $T$ by
\begin{equation}
  \label{eq:219}
  T:~~ \vt \to -\vt.
\end{equation}
Quantum mechanically $T$ is anti-unitary, i.e. accompanied by complex conjugation. Thus we get
for the basis \eqref{eq:170} and the operators \eqref{eq:174}, \eqref{eq:193} and \eqref{eq:194}
\begin{align}
  \label{eq:220}
  P: e_{b,n}(\vt) \to &~ e_{b.n}(\vt + \pi) = (-1)^n e_{b,n}(\vt), \\
  \tl{K}_0(\vt) \to & ~ \tl{K}_0(\vt+ \pi) =  \tl{K}_0(\vt), \label{eq:221} \\
  \tl{K}_1(\vt) \to & ~ \tl{K}_1(\vt+ \pi) = - \tl{K}_1(\vt),\label{eq:222} \\
  \tl{K}_2(\vt) \to & ~ \tl{K}_2(\vt+ \pi) = - \tl{K}_2(\vt), \label{eq:225}
\end{align}
and
\begin{align}
  \label{eq:223}
  T: e_{b,n}(\vt) \to &~ [e_{b.n}(-\vt)]^{\ast} =  e_{b,n}(\vt), \\
  \tl{K}_0(\vt) \to & ~ [\tl{K}_0(-\vt)]^{\ast} =  \tl{K}_0(\vt), \label{eq:224} \\
  \tl{K}_1(\vt) \to & ~ [\tl{K}_1(-\vt)]^{\ast} =  \tl{K}_1(\vt),\label{eq:40} \\
  \tl{K}_2(\vt) \to & ~ [\tl{K}_2(-\vt)]^{\ast} = - \tl{K}_2(\vt), \label{eq:226}
\end{align}

\subsubsection {A unitary transformation by a change of basis} 

In the above discussion the $b$-dependence of the representation on $H^2_{b,+}$ is contained in the Lie  operators \eqref{eq:174}-\eqref{eq:192} and in the metrical      operator
$A_b$ of Eq.\ \eqref{eq:166}, but not 
in the basis $e_n(\vt)$ of $H^2_+$ we started from. Thus, all non-equivalent irreducible representations for different $b$ are implemented by starting from
the Hardy space $H^2_+$ with the $b$-independent basis \eqref{eq:157}.
 By the  unitary transformation
\begin{equation}
  \label{eq:186}
  e_n(\vt) = e^{i\,n\,\vt} \to e_{b,n}(\vt) = e^{i\,(n+b)\,\vt}\,,
\end{equation} one can pass to $b$-dependent Hilbert spaces $\hat{H}^2_{+}(\mathbb{S}^1,d\vt;b)$ for functions with the boundary condition
\begin{equation}
  \label{eq:187}
  e_{b,n}(\vt +2\pi) = e^{2i\,b\,\pi}\,e_{b,n}(\vt)\,. 
\end{equation}
Now each irreducible unitary representation characterized by the number $b$ has its own Hilbert space,
each with the scalar product \eqref{eq:168} and with the basis
\begin{equation}
  \label{eq:189}
  \hat{e}_{b,n}(\vt) =\sqrt{\frac{(2b)_n}{n!}}\,e_{b,n}(\vt)\,.
\end{equation}
The ``reproducing kernel'' here is
\begin{align}
  \label{eq:115}
  A^{(b)}(\vt_2-\vt_1) =& \sum_{n=0}^{\infty}{e}^*_{b,n}(\vt_2)\,{e}_{b,n}(\vt_1) \\ = \nonumber
& e^{ib(\vt_1-\vt_2} [1-e^{ i\,(\vt_1-\vt_2)}]^{-2b}\nonumber \\ =& e^{ib(\vt_1-\vt_2)}\hat{A}_b(\vt_2-\vt_1)\,, \nonumber
\end{align}
The  generators \eqref{eq:174}-\eqref{eq:192} now take the form
\begin{align}
  \label{eq:188}
   \tl{K}_0 &= \frac{1}{i}\,\partial_{\vt}, \\ \tl{K}_+=& e^{i\,\vt}(\frac{1}{i}\,\partial_{\vt}+ b), \label{eq:214} \\ \tl{K}_-=& e^{-i\,\vt}\,
                                       (\frac{1}{i}\,\partial_{\vt}-b)\label{eq:215} \\ \tl{K}_1=& \cos\vt \frac{1}{i}\,\partial_{\vt} + ib\sin\vt, \label{eq:216} \\
  \tl{K}_2 &=\sin\vt \frac{1}{i}\,\partial_{\vt} - ib\cos\vt  \label{eq:217} \\ =& \tl{K}_1(\vt - \pi/2) \nonumber
\end{align}
Note that the operator $\tl{K}_2$ here, too, is obtained from $\tl{K}_1$ by replacing $\vt$ with $\vt - \pi/2$ in the latter.

Concerning the operations $P(\vt \rightarrow \vt + \pi)$ and $T (\vt \rightarrow -\vt)$ applied to the basis \eqref{eq:189} and the operators \eqref{eq:188}, \eqref{eq:216} and \eqref{eq:217} compared to
the properties \eqref{eq:220}-\eqref{eq:226} there is only a change for the basis \eqref{eq:189} for $P$:
\begin{equation}
  \label{eq:227}
 P: ~~ \hat{e}_{b,n}(\vt) \to  \hat{e}_{b,n}(\vt+\pi) =(-1)^n e^{i\pi b}\,  \hat{e}_{b,n}(\vt).
\end{equation}
The global constant phase factor $\exp (i\pi b)$ can be interpreted as representing a new type of ``fractional'' statistics in 2 dimensions \cite{kas7}, of particles
called ``anyons'' (see references below). \pagebreak[3]
\subsubsection{Aharonov-Bohm-, (fractional) quantum Hall-effects,  anyons,  Berry's phase, \\ Bloch waves etc.}
The property of the ``naive'' planar rotation operator \eqref{eq:188} to have a 1-parametric set of possible spectra - parametrized by the number $b$ - is a mathematical consequence of the fact that the operator has a 1-parametric set of self-adjoint extensions \cite{kas5}.

For physical systems topologically related to a punctured plane, the parameter $b$ can have different physical meanings \cite{kas6}:

In the description of Aharonov-Bohm effect \cite{kas6,pesh,heg} (historically more appropriate: ``Ehrenberg-Siday-Aharonov-Bohm effect'' \cite{hil}) the index $b$
is proportional to the magnetic flux $\Phi$ crossing the plane.

The magnetic flux model can also help to understand the quantum Hall effect \cite{lau1,avr,hans}. It can also do so for the fractional quantum Hall effect
\cite{lau,hans,halp}, especially in the framework of anyons \cite{kas9,rose,bart} and related Chern-Simons theories \cite{froe1,froe2}. As special Chern-Simons theories
have the structure group $SO^{\uparrow}(1,2)$ they may help to find the appropriate theoretical framework for the  dynamics associated with with the symplectic spectra
Eq.\ \eqref{eq:36}.

Closely related are the properties of Berry's phase \cite{berry1,berry2,wilc,avr}.

In the case of Bloch waves $b$ represents the momentum inside the first Brillouin zone \cite{kas8}.

\subsubsection{The operator $\tl{C}_g(K) = \tl{K}_0 + g\tl{K}_1$ on $H^2_{b,+}$}

According to Eq.\ \eqref{eq:193} the operator $\tl{C}_g(K) = \tl{K}_0 + g\,\tl{K}_1$ here has the form
\begin{equation}
  \label{eq:195}
  \tl{C}_g(K) = (1+g\cos\vt)\frac{1}{i}\partial_{\vt} + b\,(1+ge^{i\vt})
\end{equation}
The eigenvalue differential equation
\begin{equation}
  \label{eq:196}
  \tl{C}_g(K)f_{g,b}(\vt) = \tl{c}_{g,b}\,f_{g,b}(\vt)
\end{equation}
has the general solution \cite{grary}, with $0 \leq g < 1$,
\begin{align}
  \label{eq:197}
  f_{g,b}(\vt) &= C\,(1+g\cos\vt)^{-b}\,e^{i\tl{c}_{g,b}(1-g^2)^{-1/2}\,\chi(\vt)-ib\vt}, \\
               & \chi(\vt) = 2\arctan\left[\sqrt{\frac{1-g}{1+g}}\tan(\vt/2)\right], \\ \nonumber
  &  C=\text{const}. \nonumber
\end{align}
The boundary condition
\begin{equation}
  \label{eq:198}
  f_{g,b}(\vt+2\pi) = f_{g,b}(\vt)
\end{equation}
implies
\begin{equation}
  \label{eq:199}
  \tl{c}_{g,b}= (n+b)\sqrt{1-g^2}\equiv \tl{c}_{g,b;n},~n = 0,1,\ldots,
\end{equation}
The implementation of the boundary condition \eqref{eq:198} includes the transformation $\chi \to \chi + 2\pi$.

The result \eqref{eq:199}  coincides with Eq.\ \eqref{eq:91} in  Appendix~A and corresponds to the classical result \eqref{eq:93}.

Thus, we have for $\tl{C}_g(K)$ the eigenfunctions
\begin{align}
  \label{eq:272}
  f_{g,b;n}(\vt) = C\,&(1+g\cos\vt)^{-b}\,e^{i(n+b)\chi(\vt)-ib\vt}, \\ & n=0,1, \ldots, \nonumber \\
  \tl{C}_g(K)\, f_{g,b;n}(\vt) =& (n+b)\sqrt{1-g^2}\, f_{g,b;n}(\vt).
\end{align}
The constant $C$ in the solution \eqref{eq:197} can be determined from the normalization condition $(f_{g,b},f_{g,b}) = 1$. Using the relation \eqref{eq:270} leads
to the integral \cite{grad4}
\begin{align}
  1 &= |C_{g,b}|^2 \int_0^{2\pi}\frac{d\vt}{2\pi}\, (1 + g\cos\vt)^{-2b} \label{eq:271} \\
   &= |C_{g,b}|^2 (1-g^2)^{-b}\,P_{2b-1}[(1-g^2)^{-1/2}], \nonumber
\end{align}
from which the normalization constant $C_{g,b} =|C_{g,b}|$ can be determined. It is independent of $n$. Here $P_{\nu}(x),\,{\nu \in \mathbb{R}},\ x \geq 0$, is the Legendre function of the first kind
\cite{grad5,whit}. It has - among others - the properties $P_{\nu}(x) = P_{-\nu - 1}(x),\, P_{\nu}(1) =1, \, P_{\nu = 0}(x) = 1$.

\subsection{Nonlinear interactions in terms \\ of the operators $\tl{K}_j$}
The HO model plays an important role in molecular physics (see the next chapter): For example, the nuclei of diatomic molecules can oscillate relative to each other
along their connecting axis. As long as the associated energy levels are small compared to the dissociation energy $V_0$ one can approximate those vibrations by a
one-dimensional HO the potential of which is centered at the equilibrium point $q_0 = r_0 > 0$ \cite{herz2}. For higher energies when dissociation becomes relevant,
the HO is no longer an appropriate model.
\subsubsection{The Morse potential for molecular vibrations}
In order to take dissociation into account Morse suggested the potential \cite{ka14}
\begin{equation}
  \label{eq:145}
  V_{Mo}(q) = V_0\,(e^{-aq}-1)^2,\,\, V_0,\, a > 0,
\end{equation}
where $q$ is the distance of the  atomic nuclei from their point of equlibrium $q_0$.
  For $a q \ll 1$ this becomes a HO potential
  \begin{equation}
    \label{eq:146}
    V_{Mo}(q) \approx \frac{1}{2}m\omega_0^2\,q^2,\,\, \omega_0 =a\sqrt{2V_0/m},
  \end{equation} where $m$ is the reduced mass of the two nuclei.
  
  In addition
  \begin{equation}
    \label{eq:147}
    V_{Mo}(q \to \infty) = V_0, \,\,   V_{Mo}(q \to - \infty) = + \infty. 
  \end{equation}
  For $aq \ll -1$ the potential describes some kind of ``hard core''. The modifications for $q=r \geq 0$ being the radial variable are discussed in Ref.\ \cite{haar}.
  
  If $E < V_0$ the classical motions are bounded and periodic. The system is also integrable, i.e. there exist canonical angle and (constant) action variables
  in order to describe the sytem \cite{ka14}. The relationship between constant energy $E <V_0$ and action variable $I$ turns out to be (see Eq.\ \eqref{eq:278})
  \begin{equation}
    \label{eq:148}
    \omega_0I = 2V_0(1-\sqrt{1-E/V_0},\;\; E= \omega_0I\,\left(1-\frac{\omega_0I}{4V_0}\right).
  \end{equation}
  This gives the Hamilton function
   \begin{equation}
    \label{eq:149}
    H_{Mo}(\vp, I) = \omega_0 I\left(1-\frac{\omega_0I}{4\,V_0}\right),
  \end{equation}
  with the associated Eqs.\ of motion
  \begin{align}
    \label{eq:238}
    \dot{\vp} =& \partial_IH = \omega_0(1-\frac{\omega_0I}{2V_0}) \equiv \omega_I, \\
    \dot{I} =& - \partial_{\vp}H = 0,~ \Rightarrow\; \omega_I = \text{ const.}  \label{eq:239}
  \end{align}
  Note that here
  \begin{equation}
    \label{eq:250}
    E(I) = \frac{1}{2}(\omega_0 + \omega_I)\,I.
  \end{equation}
  Replacing the action variable $I$ in the Hamilton function \eqref{eq:149}
  by the operator $\hbar \tl{K}_0$ leads to the spectrum
  \begin{align}\label{eq:150}
    E_{b,n}=& \hbar\omega_0(n+b)\left[1-\frac{\hbar\omega_0}{4\,V_o}(n+b) \right], \\
    E_{b,n=0}=& \hbar\omega_0b\left(1-\frac{\hbar\omega_0}{4\,V_o}b\right). \label{eq:243}
  \end{align} For the bracket [\ldots] in Eq.\ \eqref{eq:150} to be positive only those $n$ are allowed which imply this property.

  This system is a simple but instructive example how the use of the canonical pair angle and action variables instead of the canonical position and momentum can simplify the description
  of the dynamics of the system, at the expense of making it intuitively less accessible! That might be especially so if the system is not completely integrable and
  the action variable a function of time, too, as in the model of Ch.\ II above.
  \subsubsection{Potentials involving the terms $\tl{K}_+\tl{K}_-$ and $\tl{K}_-\tl{K}_+$}
Due to the Casimir operator relations \eqref{eq:65} the eigenvalue equations of the (dimensionless) Hamiltonians (up to a factor $\hbar \omega$)
\begin{equation}
  \label{eq:100}
  \tl{F}_- = \tl{K}_0 +g_-\,\tl{K}_+\tl{K}_-
\end{equation}
and
\begin{equation}
  \label{eq:101}
   \tl{F}_+ = \tl{K}_0 +g_+\,\tl{K}_-\tl{K}_+
 \end{equation} can be solved immediately:
 Eigenvectors are still those of $\tl{K}_0$ (see Eq.\ \eqref{eq:36}) and the eigenvalues are
 \begin{equation}
   \label{eq:102}
   \tl{f}_{g_-;b,n} = n+b+ g_-(n+2b-1)n\,;~ \tl{f}_{g_-;b,n=0} = b,
 \end{equation}
 and
 \begin{equation}
   \label{eq:103}
    \tl{f}_{g_+;b,n} = n+b+ g_+(n+2b)(n+1)\,;~ \tl{f}_{g_+;b,n=0} = b(1 + 2g_+).
  \end{equation}
  The models describe the annihilation and creation of quanta (Eq.\ \eqref{eq:100}) and vice versa (Eq.\ \eqref{eq:101}). The couplings $g_-$ and $g_+$ may depend on external parameters.
 \section{Reflections on  possible   experiments \\ and observations}
 Replacing the ingrained and very successful habit of describing the quantum HO by the ``canonical'' pair position and momentum operators (or the associated creation and annihilation operators) by the quantum version of its classical -  locally - equivalent canonical pair angle and action variables may appear unnecessarily artificial and even unnatural:

 Compared to position and momentum variables  the pair angle and action variables is less  familiar as far as
 visualization and perception are concerned:

 Whereas the angle can be illustrated well as a fraction of the unit circle and its s-fold coverings by the
   corresponding number of rotations of the hand of a clock, a visualization of the action variable is not so obvious. True, all quantum mechanical action variables must - in principle - be proportional to Planck's constant  $\hbar$ and  for integrable systems  it appears to be closely related to the conserved quantity ``energy''.
   But we have seen in Ch.\ II that the action variable $I$ may be quite useful as a coordinate even if it is not a constant of motion. For such time--dependent
   $I(t)$ the relation \eqref{eq:285} may be a helpful tool for an intuitive interpretation.

   An important lesson from Ch.\ II for the discussions below is that the energy $E$ may be conserved even if the action variable $I(t)$ is not!

   Perhaps we have to go beyond the use of position and momentum as the basic observables in a part of the quantum world where other  ``canonical'' observables are more appropriate! This is, of course, a larger challenge for a reformulation of (perturbative) quantum field theories etc., for which the orthodox description of the HO is a fundamental building block!

   In view of the qualitative  differences between the global phase spaces \eqref{eq:7} and \eqref{eq:8} and their possible physical implications - especially for the associated quantum theory - it  is obviously important to make experimental and observational attempts to look for corresponding phenomena in nature!

   All the following considerations apply, of course, only, {\em if}  the mathematical models from the previous chapters have counterparts in nature! For this reason all possible applications discussed in the following are  hypothetical! The good news is that the relevance of the model can be tested in the laboratory and by astrophysical
   observations! The (hopefully preliminary) bad news is that the associated theoretical framework for the {\em dynamics} governing  transition rates etc. involving the new spectra  still has to be worked out!
 \subsection{Generalities}
 In view of their possible far-reaching implications the above theoretical results should, of course, be subject to critical reviews and be probed experimentally!
 In the following - as a kind of ``tour d'horizon'' -  ideas and
suggestions for such  experiments and observations are discussed, in the hope that a few experimentalists will be motivated and inclined to  meet the challenge and that experts - experimentalists and theoreticians - in the areas of physics mentioned below, will point out possible misunderstandings  and will suggest improvements and consequences!

Harmonically oscillating quantum systems can be found in many areas of physics, at least approximately close to the corresponding (local) minima of  classical ``binding'' potentials with periodic motions.

It is important to note that the ``symplectic'' or ``fractional'' spectrum \eqref{eq:31} is tied to the groups $U(1)$ or $O(2)$ and their infinitely many coverung groups, but not to the rotation group $SO(3)$ and its
  single  2-fold  covering $SU(2)$. Accordingly one has to look for 2-dimensional (sub)systems with ``effective'' phase spaces  \eqref{eq:8}. Such systems may be found in
  molecular spectroscopy (e.g.\ diatomic molecules), quantum optics, optomechanics and - possibly - in astrophysics (``dark'' energy and ``dark'' matter, see below).

  One obvious question is: Why haven't we seen those symplectic $b$-dependent spectra  yet? Several answers are possible:
  
  0. They just don't exist in nature!
  
  1. One possible reason is that no one has looked for them. This is quite plausible if the ``visibility'' of those symplectic spectra is very weak, as, e.g.\ for infrared  emission or absorption lines of homonuclear diatomic molecules like H$_2$, because they have no electric dipole moment or because their Stokes or Anti-Stokes lines  in inelastic Raman scattering off vibrating and rotating molecules (see below) are very weak.

  2. As discussed in Section III.C the impact of the (composite) "orthodox" Fock space annihilation and creation operators \eqref{eq:51} and \eqref{eq:231} with the usual
  properties
  \eqref{eq:53}, \eqref{eq:230} and \eqref{eq:54} may dominate and obscure the symplectic spectra \eqref{eq:41}, except for the value $b=1/2$. Thus, one has to find means in order to
  discover other (fractional) parts of the spectra \eqref{eq:41}, if they exist at all! In any case,  their observability appears to be rather weak.

 3.  Transitions - radiative, non-radiative, collisional, Raman-type etc.  - between different levels of the spectra \eqref{eq:41} require  appropiate kinds of electromagnetic interactions, the dynamics of which has not yet been worked out!

  Consider two generally different levels of the spectra \eqref{eq:41}:
  \begin{equation}
    \label{eq:240}
    E_{b_j,n_{b_j}} = \hbar\omega_j(n_{b_j} + b_j),\,j=1,2;\,n_{b_j} = 0,1,\ldots.
  \end{equation}

  For a fixed $b= b_1=b_2;\,$ and $\omega = \omega_1 = \omega_2$ the observable energy difference between an upper level characterized by $n_b'$ and a lower level characterized by $n_b'' < n_b'$.
  \begin{equation}
    \label{eq:232}
    E_{b,n_{b}'} -E_{b,n_{b}''} = \hbar \omega (n_b' -n_b'')
  \end{equation}
  cannot be distinguished from the corresponding difference for a $b_2 \neq b$.

  More interesting is a transition with a change of the B-index
  ($b_1 \leftrightarrow b_2$): 
  \begin{equation}
    \label{eq:233}
     E_{b_1,n_{b_1}'} -E_{b_2,n_{b_2}''} = \hbar \omega [ (n_{b_1}' -n_{b_2}'') + (b_1-b_2)].
  \end{equation}
  If such transitions are possible, e.g.\ for $n_{b_2}'' = n_{b_1}' =0$ and as - up to now - the only condition on  $b_2$ (and $b_1$) is the inequality $b_2>0$ one may have a cascade of (fluorescence) transitions
  \begin{equation}
    \label{eq:234}
    b_1 \rightarrow b_3 \rightarrow \cdots b_m \rightarrow b_2 > 0,
  \end{equation}
   accompanied by the emission of $m-1$ low-frequency (lower than $\omega$) quanta. Even a continuum
  between $b_1$ and $b_2$ appears possible. All this depends on the still to be established associated dynamics, which  determines rates and selection rules!
  If the initial quanta cascade down the ``fluorescence'' sequence \eqref{eq:234} they can end up in the microwave or even radiowave region, without loss of the total energy!
  \begin{figure}[h]\includegraphics[width=\columnwidth]{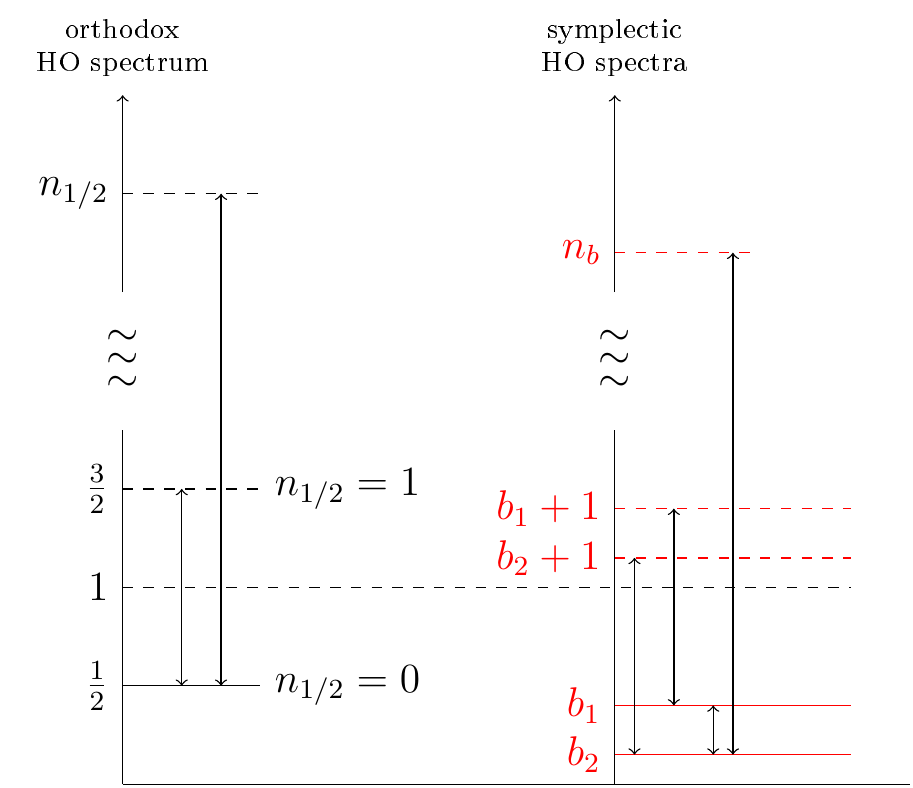}\caption{\label{fig:newplot3-1} Comparison of the orthodox and the symplectic spectra of the HO,
      with indications of possible transitions between levels (ignoring any  selection rules) }
\end{figure}
  
If $\omega_2 \neq \omega_1$ (occurs for diatomic molecules with different isotopic atoms and for electronic transitions between local minima of different Born-Oppenheimer potentials for the nuclei; see below), one has
\begin{equation}
  \label{eq:273}
  (E_{b_1,n_{b_1}'}-E_{b_2,n_{b_2}''})/\hbar = \omega_1[n_{b_1}' - \sigma\, n_{b_1}'' + b_1 - \sigma b_2],~ \sigma = \frac{\omega_2}{\omega_1}.
\end{equation}

4.  As discussed in Section III.B above, for a given $\omega$ one needs the time $t_s, \,\omega\,t_s = 2\pi s\,, \, s \in \mathbb{Z},$ in order to ``run'' through an $s$-fold
covering $\mathbb{S}^1_{[s]}$ of the circle $\mathbb{S}^1$. The prefactor $\exp(-ib\, \omega t)$ of the general state \eqref{eq:208} shows that here the time  ``angle'' $\tl{t} = \omega t$ can be reduced by a small $b\ll 1/2$!

 For infrared light the number $s$ of coverings is obviously very large for a finite time interval
$\Delta t \gg T = 1/(2\pi \omega)$, where $\omega \approx 10^{14}$Hz in the near infrared.

5.   In Section III.D above we discussed the $n \to n\pm 1$ transition amplitudes for the ``primitive'' effective Hamiltonian \eqref{eq:69} the interaction term of which
  mimics partial properties of  an electric dipole moment. If one wants to include the influence of external electromagnetic radiation, one has to allow the coupling term
  $g$ to depend explicitly on time or via other external parameters or fields \cite{merz}.

6.   For experimental tests it is essential to find quantities (``observables'') which are especially sensitive to values of the B(argmann)-index $b$. The following is an - incomplete -
 list of possible theoretically interesting experiments (without proper knowledge of their feasibility in the laboratory or of their observability in the sky)!

 \subsection{Vibrating diatomic molecules}
 Among the most important and interesting oscillators the above results may apply to are vibrating diatomic molecules (for introductions to their physics see, e.g.\ the textbooks
 \cite{herz,land2,gasi,atk,atk2,pari}). They have one vibrational degree of freedom: oscillations  about the point of equilibrium along the line connecting the two nuclei (``internuclear axis'' = INA). Near that equilibrium point the potential
 may be considered to be harmonic. In the Born-Oppenheimer (BO) approximation the effective potentials for the vibrating nuclei are provided by  energy configurations of the electron ``cloud'' the dynamics of
 which depends only ``adiabatically'' on the state of the nuclei, especially on their distance $R$ (see Fig.4).

 The (classical) angular frequency $\omega = 2\pi\nu$ for the mutual harmonic vibrations of the nuclei
 is given by
 \begin{equation}
   \label{eq:235}
   \omega = 2\pi\,\nu = \sqrt{k/\mu},
 \end{equation}
 where $k$ is the ``force constant'', determined - in the BO approximation - solely by the actual electronic configuration and $\mu = m_1\, m_2/(m_1 + m_2)$ is the ``reduced''
 mass of the two vibrating atoms.

 Spectroscopists denote the vibrational level numbers $n$ of the HO by $v$ and give the frequencies $\nu\,$[s$^{-1}$] in terms of the
 ``wave number'' $\tl{\nu} = \nu/c\,$[cm$^{-1}$]. One then has the (approximate) equivalences
 \begin{equation}
   \label{eq:279}
    1\,\text{eV}  \cong 8066\, \text{cm}^{-1}\cong 11605^{\circ} \text{ K}.
 \end{equation}
 Spectroscopically  the differences between homonuclear (equal nuclei like molecular hydrogen H$_2$ or oxygen O$_2$) and heteronuclear (different nuclei like carbon monoxyde $^{12}$C$^{16}$O) diatomic molecules are important: because of space reflection symmetry the homonuclear molecules do not have a permanent electric dipole moment and, therefore, no corresponding infrared emissions or absorptions \cite{land3}. If, however, their polarizability is nonvanishing, they can have induced electric dipole moments, e.g. in case of elastic and inelastic Raman-type scattering or by collisions.

 In addition to the vibrational energy levels characterized by the numbers $v=0,1,2 \ldots$ the diatomic molecules have rotational levels $J=0.1.2, \ldots$ due to the
 rotations of the molecule around an axis which lies in a plane perpendicular to the INA and passing through the centre of mass on that axis. So in general one has the combined vibration - rotation (``rovibrational'') transitions $(v',J') \to (v'',J'')$. The frequencies of the vibrational transitions are generally in the ``near-infrared'' (frequencies around $\nu \approx 10^{14}$\,s$^{-1}$) and those of the  rotational ones are at least one order of magnitude smaller and are in the ``far-infrared'' or microwave region.

 Example: Molecular hydrogen H$_2$

 Here are some essential properties of the molecule H$_2$ which are importent for our present discussion: As a homonuclear diatomic molecule H$_2$ has no permanent
 electric dipole moment (this property is frequently mentioned in the literature, but very rarely proven; for a proof see Ref.\ \cite{land3}). Because of this missing electric dipole moment their is no corresponding infrared emission or absorption.

\begin{figure}[h]\includegraphics[width=\columnwidth]{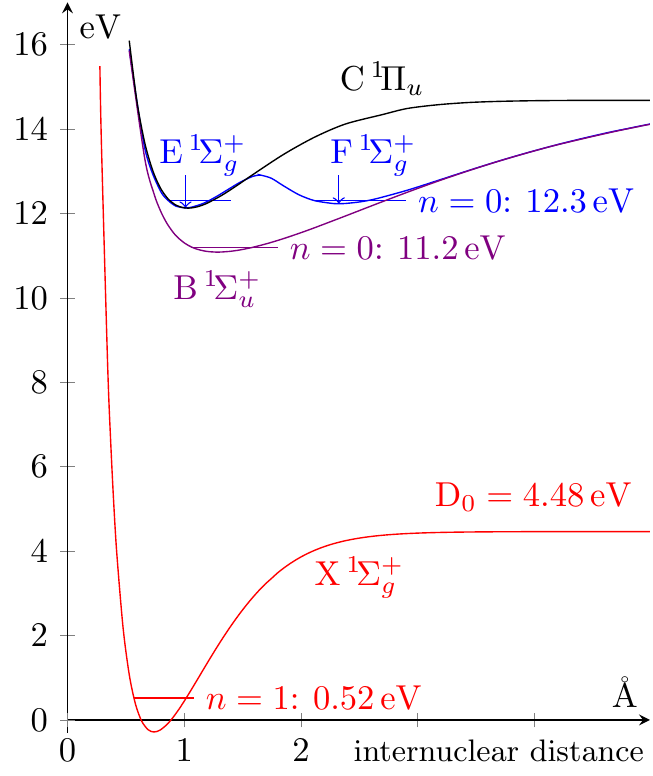}\caption{\label{fig:potentialh2.pdf} Some of the lowest "binding" electronic BO potentials for the
     vibrations of the two H$_2$ nuclei as a function of their distance $R$. The zero energy on the ordinate coincides with the zero-point energy of
     the electronic potential X $ ^1\Sigma^+_g$. The numerical values are taken  from Ref.\ \cite{sha}. For the interpretation of the terms
   denoting the different curves see Appendix~B.}
\end{figure}
 There is, however, (weak) magnetic dipole and electric quadrupole infrared radiation
 \cite{rou}.
 
 Due to that missing electric dipole moment there are no direct vibrational transitions $v \leftrightarrow v \pm 1$ within a given electronic BO -- potential, like the
 electronic ground state potential X $^1\Sigma^+_g$ (see Fig.\ 4).

 As a consequence, in order to experimentally analyse the ladder of vibrational states of, e.g.\ the BO electronic ground state potential
 X $^1\Sigma^+_g$,  an ``ultraviolet detour'' has to be taken: one first initiates an  ultraviolet allowed (1- or 2-$\gamma$) absorptive transition from the electronic ground state to a vibrational level of
 a higher BO electronic potential (Fig.\ 4), from where the photons cascade down (in 1 or more steps) to a  vibrational level of X $^1\Sigma^+_g$ which is different of
 the one the photons originally started from. The difference of the observed ultraviolet frequencies then provides information about the vibrational levels of the selected BO potential
 \cite{sha,glass,ster,abg1,abg2,ron,bail,han,niki,dic,niu,alt}.

 The ``ultraviolet detour'' also plays an essential role in the so-called ``Solomon process'' which leads to photodissociation of H$_2$ \cite{field,stech,loeb5,loeb}.

 Another possibility to observe vibrational and rovibrational levels of  H$_2$ electronic BO potentials is provided by the polarizability of the molecule, which
 allows for Raman-type transitions associated with induced electric dipole moments, induced by by external light beams or be collisions\cite{polar,veirs,mcca,long,li}.
 
The two nuclei (protons) oscillating in the binding electronic BO potentials may have antiparallel spins (para-H$_2$) or parallel ones (ortho-H$_2$).
 For recent
 summeries and reviews of the role of H$_2$ in different areas of physics see, e.g.\ \cite{field,sha,spre,uba,wak,hoel}. More  references will be quoted in the course of the discussions below. (Numerical values of  quantities mentined below are rounded up/down from their impressively   determined accurate theoretical and experimental values).
 
 For the nuclear vibrations of the diatomic homonuclear molecules H$_2$ in the electronic ground state X $ ^1\Sigma^+_g$ BO potential  (Fig.\ 4) one
 has for the ``transition'' (''ground tone'')  \cite{kom,dic}
 \begin{align}
   \label{eq:241}
  (\Delta\,v, \Delta\,J):\,& (v''= 0, J''=0) \leftrightarrow (v'= 1, J'=0) \\ &\approx (\pm [\tl{\nu}_{0\leftrightarrow 1}  \approx 4161 \text{ cm}^{-1}],0) \nonumber
 \end{align}
 which is one of the larger values for vibrating diatomic molecules.

 Recall that the BO electronic ground state X$ ^1\Sigma^+_g$ is an effective potential for the vibrations
 of the two nuclei, depending on their distance $R$ (Fig.\ 4).
 
 The vibrational transition value \eqref{eq:241} correponds to about $0.516 \,$ eV, a wave length $\lambda =1/\tl{\nu} \approx 2,4\, \mu$m and a temperature of $\approx 5988^{\circ} $ K. 

 In comparison the rotational transition $(0,0) \leftrightarrow (0,1)$ has the wave number $\tl{\nu} \approx 118$ cm$^{-1} \cong 0.0146$ eV \cite{kom}.  This  means 
 a wavelength $\lambda   \approx 85\,\mu$m.

 If the vibrating H$_2$ molecule were an ideal HO, its ``orthodox'' zero-point energy, according to Eq.\ \eqref{eq:241}, would be
 \begin{align}
   \label{eq:242}
   E_0(X~  ^1\Sigma^+_g)& = \frac{1}{2}\tl{\nu}_{0 \leftrightarrow 1}(X~  ^1\Sigma^+_g) \\ & \approx 2080 \text{ cm}^{-1} \cong 0.258 \text{ eV} \cong 2994^{\circ} \text{ K}.
\nonumber  \end{align}

 The vibrating molecule H$_2$ is, of course, no ideal HO because it dissociates at a finite energy $D_0> E_0$. The Morse potential \eqref{eq:145} takes this qualitatively into account,
 as can be
 seen from the relations \eqref{eq:147}. The ``anharmonic'' modifications of energy \eqref{eq:149} and angular frequency \eqref{eq:238} are small as long as
 $\omega_0I \ll V_0$.

 The quantum mechanical energy \eqref{eq:150} can be written as
 \begin{equation}
   \label{eq:244}
    E_{b,n}= \hbar\omega_0\,(n+b)-\frac{(\hbar\omega_0)^2}{4\,V_o}(n+b)^2,
 \end{equation}
 which may be considered as a polynomial in $(n+b)$.  As the Morse potential still is only a rough approximation, one has taken - for the orthodox value $b=1/2$ -
 the expression \eqref{eq:244} as a suggestion to parametrize the vibration and rotation levels generally by \cite{dun,herz2,iri}
 \begin{equation}
   \label{eq:245}
   E_{v ,J } = \sum_{i\ge 0, j\ge 0}Y_{i,j}\, (v+1/2)^iJ^j(J+1)^j,
 \end{equation}
 where the coefficients $Y_{i,j}$ are determined (mainly) experimentally. The ground state (``zero-point'') energy is given by
 \begin{equation}
   \label{eq:246}
   E_{0,0} = \sum_{i \ge 0}Y_{i,0}(1/2)^i.
 \end{equation}
 For the Morse potential one has $Y_{1,0}= \hbar \omega_0,\,Y_{2,0} = - \hbar \omega_0/(4V_0)$, all other $Y_{i,0}$ vanishing.

 The approximation ansatz \eqref{eq:245} gives for H$_2$ instead of \eqref{eq:242} the  value \cite{iri}
 \begin{equation}
   \label{eq:247}
   E_0[\text{H}_2) \approx 2180 \text{ cm}^{-1}.
 \end{equation}
 Thus, {\em by passing from the orthodox HO spectrum ($b= 1/2$), usually  associated with H$_2$ infrared vibrations, to the symplectic one [$b \in (0,1)$] one can lower the zero-point  energy of the BO potential X~$^1\Sigma^+_g$ maximally by the (approximate) amount}
 \begin{align}
   \label{eq:248}
 E_0(H_2;& b\to 0) - E_0(H_2; b=1/2) \equiv  \hat{E}_0(H_2) \\   \approx & 2100 \text{ cm}^{-1} \cong 0.26 \text{ eV} \cong 3000^{\circ}\text{ K }. \nonumber 
\end{align}
Similarly, the known dissociation energy of H$_2$ \cite{spre,che,hoel,puch1}
\begin{equation}
  \label{eq:249}
  D_0 \approx 36118 \text{ cm}^{-1} \cong 4.48 \text{ eV}
\end{equation}
- theoretically - increases maximally by the the amount \eqref{eq:248}.

Thus, the orthodox H$_2$ vibrational spectrum (b = 1/2) can be considerably ``detuned'' for $1/2 > b > 0$.

The difference \eqref{eq:248} between the orthodox and the symplectic ground states of the vibrating H$_2$ molecule implies an additional effecive Boltzmann factor
\begin{equation}
  \label{eq:255}
 e^{-\hat{E}_0(H_2)/(k_B T)} = e^{-3000^{\circ}/T }.
\end{equation}
Preliminarily ignoring {\it all} dynamical mechanisms the last Eq.\ says that for $T < 3000^{\circ}$K the symplectic ground state becomes preferred {\it statistically}.
This will play a role in our astrophysical discussion below. It also indicates that the symplectic HO spectra may be observed better at very low temperatures.

As mentioned above, in the BO approximation  the electronic ground state X $ ^1\Sigma^+_g$ (which includes the action of the nuclear Coulomb potentials on the electrons)  provides a potential for the two vibrating nuclei as a function of their distance $R$. The potential has a minimum around which the oscillations are approximately harmonic. The same applies to the next higher electronic (metastable) states B $^1\Sigma_u^+$, E\,F $ ^1\Sigma^+_g$ and C $^1\Pi_u$. They have local minima in appropriate neighbourhoods of which the vibrations are harmonic, too (see Fig. 4).

In the following list one can find the measured energy differences between the ground states of the different electronic levels relative to X $ ^1\Sigma^+_g(v=0,J=0)$ and the energies $E_{0 \leftrightarrow 1}$ of the first vibrational excitations $(v=0,J=0) \to (v=1, J=0)$ above those ground states  \cite{sha,glass,ron,abg1,abg2,hann,bail,niki,niu}. The  data here are from Ref.\ \cite{bail}:
\begin{align}
  \geq & \text{X}\, ^1\Sigma^+_g\,[\text{eV}] & E_{0 \leftrightarrow 1}\,[\text{eV}] \label{eq:251} \\
 \text{X}\, ^1\Sigma^+_g ~~~~~&  0 & 0.52~~~~~ \nonumber \\
  \text{B}\, ^1\Sigma_u^+~~~~~ &11.19 & 0.16 ~~~~~ \nonumber \\
  \text{E}\,  ^1\Sigma^+_g~~~~~ &12.30& 0.29 ~~~~~ \nonumber \\
  \text{F}\,  ^1\Sigma^+_g~~~~~ &12.32 & 0.15 ~~~~~ \nonumber \\
  \text{C}\, ^1\Pi_u~~~~~ &12.30 & 0.29 ~~~~~ \nonumber
\end{align}
The second $E_{0 \leftrightarrow 1}$-column shows that the first vibrational excitations are generally quite different for the different electronic levels, reflecting
the curvature differences at the minima of the potential curves. If the five BO potentials are approximately harmonic near their minima, the above numerical values of
$E_{0 \leftrightarrow 1}$ are {\em twice} the values of  their zero-point energies.

Note that the transitions from (to) the listed higher {\it electronic} levels to (from) the electronic ground state X $ ^1\Sigma^+_g(v=0,J=0)$  are in the vacuum UV ($ \geq$ 6.20 eV). They are approximately the same as the Lyman $\alpha$ transition of atomic hydrogen (10.20 eV). This is important for a gas mixture of H and H$_2$: The relative  energy differences \eqref{eq:251} are all larger than the Lyman $\alpha$ transition and they become even larger for the symplectic spectra. This is important for astrophysical applications (see below).

As mentioned above, in order to determine the vibrational transition energies $ E_{0 \leftrightarrow 1}$ in the list \eqref{eq:251} experimentally one has to take a ``detour'' of determining related (electronic) UV transitions first and then subtract the corresponding energies \cite{spre,dic,alt,che,hoel}.

The set of UV transitions from the states of the potential $ \text{X}\, ^1\Sigma^+_g $ to those of $ \text{B}\, ^1\Sigma_u^+$ or vice versa is called the ``Lyman-band'', and those of
$ \text{X}\, ^1\Sigma_g^+$ to  $\text{C}\, ^1\Pi_u$ the ``Werner-band'' \cite{glass,ron,abg1,abg2,bail}. \pagebreak[1]
\subsection{ Vibrations of diatomic molecules \\  with different isotopic atoms}
Such systems played an important but nowadays mostly forgotten role in the early history of quantum mechanics:

Even before Heisenberg derived the now well-established spectrum of the HO in his famous paper from July 1925 \cite{heis}, Mullikan had concluded from his
investigations 
of diatomic molecules that their vibrational spectra should be described by
half-integers, not integers as the Bohr-Sommerfeld quantization prescription had suggested \cite{born,born1}. Mullikan compared  the vibrational spectra of diatomic molecules in which
one atom was replaced by an isotope (B$^{10}$O and B$^{11}$O; AgCl$^{35}$ and AgCl$^{37}$) \cite{mull}.

Classically the vibrating atoms have angular frequences  $\omega_i = \sqrt{k/\mu_i}$,\, where the 
$\mu_i, i=1,2$ denote the reduced masses of the oscillators, $\mu_1$ for one and $\mu_2$ for the other molecule containing one or two isotopic atoms. The (electronic) oscillator strength $k$ is assumed to be the same in both cases (BO approximation).

Let $E_0(i) = \gamma\,\hbar \omega_i$ be the two slightly different oscillator ground state energy levels for the two ``isotopic''
oscillators. Let further $E_a$ and $E_b$ be two known electronic energy levels (they may be equal) from which transitions to the ground states with energies
$E_0(i)$ are possible. Then the difference
\begin{equation}
  \label{eq:98}
  \omega_{a,1} - \omega_{b,2} = (E_a-E_b)/\hbar - \gamma (\omega_1-\omega_2)
\end{equation}
of the frequencies
\begin{equation}
  \label{eq:99}
  \omega_{a,1} =[E_a-E_0(1)]/\hbar,\;\; \omega_{b,2}= [E_b-E_0(2)]/\hbar
\end{equation}
can be used in order to determine $\gamma$. Mullikan concluded that $\gamma \approx 1/2$. A good review of the method can be found in Ref. \cite{herz3}.

Due to the tremendous experimental and technological advances since those experiments from almost 100 years ago it appears possible to perform similar more refined experiments \cite{dic} in order to find 
fractional values of the B-index $b$ other than $1/2$. However, one first has to account for the deficits of the BO approximation and for the corrections due to rotational, electronic and QED effects \cite{pach2,puch1}!

\subsection{Interferences of time dependent energy eigenstates}
  Applying the unitary time evolution operator \eqref{eq:43} to the energy eigenstates $|b,n\rangle$ yields ($\hbar =1$ in the following)
  \begin{equation}
    \label{eq:111}
    U(\tl{t})\,|n,b\rangle = e^{-i(n+b)\tl{t}}|n,b\rangle,\;\tl{t}= \omega\,t.
  \end{equation}
  Reccall that $\tl{t}$ is a (dimensionless) angle variable.
  Let $\tl{t}$ increase by an amount $\delta\,\tl{t}$ which may be implemented by either a change of $\omega$ or of $t$ or of both.
  Consider the superposition
  \begin{equation}
    \label{eq:112}
    |n,b;\tl{t},\delta\tl{t}\rangle = (1+e^{-i(n+b)\delta \tl{t}})e^{-i(n+b)\tl{t}}|n,b\rangle.
  \end{equation}
  Then the oscillations of the ``intensity''
  \begin{equation}
    \label{eq:113}
    |\langle n,b;\tl{t},\delta\tl{t}|n,b;\tl{t},\delta\tl{t}\rangle|^2 = 4\cos^2[(n+b)\delta\tl{t}/2].
  \end{equation}
  are sensitive to the value of $b\delta\tl{t}$, especially for $n=0$. For an analogous approach in a recent experiment see Ref.\ \cite{mcc}.

  An   alternative to generate such interferences by a change $\delta \tl{t}$ one may also use - at least theoretically - a change $\delta b$. The question, how to
  generate states like $|n,b\rangle$ experimentally  has, unfortunately, to be left open here.

  \subsection{Transitions associated with the Hamiltonian  $H(K) = \hbar\,\omega\,\tl{C}_g(K)$}
  In case the model Hamiltonian \eqref{eq:69} with its "effective" electric dipole moment  can somehow be implemented experimentally, either by heteronuclear molecules
  like. e.g.\ $^7$LiH (it has the rather large electric dipole moment 5.9 D[ebeye]) or by Raman-type induced electric dipole moments of homonuclear diatomic molecules,   then especially the transitions \eqref{eq:73} depend sensitively on the value of $b$: The probability for the transition $ n=0 \leftrightarrow n=1$ is given by 
  \begin{equation}
    \label{eq:281}
     |\langle b, m=0|\tl{C}_g(K)|b, n=1\rangle|^2 = b\,g^2/2. 
  \end{equation}
   So, if the index $b$ is very small $>0$ - as it appears to be in astrophysical cases (see below) - then the same holds for that transition probability!
  
  Another essential point here is  that the spectrum $\{n+b\}$ of $\tl{K}_0$ is rescaled for $\tl{C}_g(K)$ by an overall  "redshifting" factor $\sqrt{1-g^2}$
  (see Eq.\ \eqref{eq:272}).
  \subsection{Traps for neutral molecules and optomechanics}
  A speculatively optimal experimental situation would be a diatomic neutral molecule in a cooled down trap which allows the vibrational emission
  and absorption  properties of the molecule to be observed, especially those of its different electronic potential ground states. As already stressed above, the conditions are different  for heteronuclear and homonuclear molecules, the former having an electric dipole moment, the latter not, which requires some Raman-type excitations.
  In view of the very impressive developments of experimental
  possibilities involving such traps \cite{gri,leib,ash}, it appears possible to achieve at least a few of the required aims. Closely related are optical devices
  coupled to mechanical oscillators \cite{asp,mcc,fors,qiu}
  
  \subsection{Perelomov coherent states}
  Among the three different types of coherent states \cite{kas10} associated with the Lie algebra \eqref{eq:105},  the so-called ``Perelomov'' coherent states
  appear to be the most promising ones in order to detect traces of HO spectra with $b\neq 1/2$: Their matrix elements contain the Bargmann index $b$ quite explicitly
  and they can be generated experimentally \cite{kas10}.

  The states $|b, \lambda\rangle$ can either be defined as eigenstates of a composite ``annihilation'' operator,
  \begin{align}
    E_{b,-}|b,\lambda\rangle =& \lambda\,|b,\lambda\rangle,\;\; E_{b,-} = (\tl{K}_0+b)^{-1}\tl{K}_{-},\label{eq:116} \\
    & \lambda = |\lambda|e^{-i\theta} \in \mathbb{D} = \{\lambda \in \mathbb{C},\; |\lambda| < 1\}, \nonumber
  \end{align}
  or by generating them from the ground state $|b,0\rangle$ by means of the unitary operator
  \begin{align}
    U(\lambda)_P  = & e^{(w/2)\tl{K}_+ - (w^{\ast}/2)\tl{K}_-} = e^{\lambda\,\tl{K}_+}e^{\ln(1-|\lambda|^2)\tl{K}_0}e^{-\lambda^{\ast}\tl{K}_-}, \label{eq:117} \\
    & w = |w|e^{-i\theta} \in \mathbb{C},\;\; \lambda = \tanh(|w|/2)e^{-i\theta}, \nonumber \\ &|w| = \ln\left(\frac{1+|\lambda|}{1-|\lambda|}\right), \nonumber
  \end{align}
  so that
  \begin{equation}
    \label{eq:118}
    |b,\lambda\rangle = U(\lambda)_P|b,0\rangle.
  \end{equation}
  In terms of number states we have the expansion
  \begin{align}
    \label{eq:119}
    |b,\lambda\rangle &=  (1-|\lambda|^2)^b\sum_0^{\infty}\left(\frac{(2b)_n}{n!}\right)^{1/2}\lambda^n|b,n\rangle, \\
     (2b)_n & = 2b(2b+1) \ldots  (2b+n-1) = \Gamma(2b+n)/\Gamma(2b), \nonumber \\ &~~~~ (2b)_{n=0}=1. \nonumber
  \end{align}
  Note that the coefficient of $\lambda^n$ in this expansion is the same as that of $e_n(\vt)$ in Eq.\ \eqref{eq:170}.
  
  Important expectation values with respect to $|b,\lambda\rangle$ are
  \begin{align}
    \langle b,\lambda|\tl{K}_0|b,\lambda \rangle \equiv & \overline{\tl{K}_{0}}_{;b,\lambda} = b\cosh|w|, \label{eq:120} \\
    \langle b,\lambda|\tl{K}_1|b,\lambda \rangle \equiv & \overline{\tl{K}_1}_{;b,\lambda} = b\sinh|w|\,\cos\theta,\label{eq:121}, \\
    \langle b,\lambda|\tl{K}_2|b,\lambda \rangle \equiv & \overline{\tl{K}_2}_{;b,\lambda} = - b\sinh|w|\,\sin\theta, \label{eq:122}\\
    \langle b,\lambda|N|b,\lambda \rangle \equiv & \overline{N}_{b,\lambda} = b(\cosh|w|-1), \label{eq:123} \\
    (\Delta N)^2_{b,\lambda} \equiv & \overline{N^2}_{b,\lambda}- \overline{N}_{b,\lambda}^2 = \frac{1}{2}b\sinh^2|w| \label{eq:124}
  \end{align}
  It follows that most quantities can be expressed in terms of the ``observables'' $\overline{N}_{b,\lambda}$ and $(\Delta N)_{b,\lambda}$:
  As
  \begin{equation}
    \label{eq:125}
    b\sinh|w| = \sqrt{2b}(\Delta N)_{b,\lambda}
  \end{equation}
  and $\cosh|w| = (\sinh^2|w|+1)^{1/2}$ we have, e.g.,
  \begin{align}
    \overline{\tl{K}_{0}}_{;b,\lambda} =& [2b(\Delta)^2_{b,\lambda}+b^2]^{1/2}, \label{eq:126} \\
    \overline{N}_{b,\lambda}[1+\overline{N}_{b,\lambda}/(2b)] =& (\Delta N)^2_{b,\lambda}. \label{eq:127}                                                                
  \end{align}
  It follows from the last equation that Paul's parameter R \cite{pau} here has the value
  \begin{equation}
    \label{eq:128}
    R_{b,\lambda} \equiv \frac{(\Delta N)^2_{b,\lambda}-\overline{N}_{b,\lambda}}{\overline{N}_{b,\lambda}^2} =\frac{1}{2b}.
  \end{equation}
  $R$ is a measure for deviations from a Poisson distribution for which $(\Delta n)^2 -\bar{n} = 0$.

  In addition we have
  \begin{align}
    \label{eq:129}
    |\lambda| = & \tanh(|w|/2) =\frac{\cosh|w| -1}{\sinh|w|} = \frac{\overline{N}_{b,\lambda}}{\sqrt{2b}(\Delta N)_{b,\lambda}} \\
    |\lambda|^2 = & \frac{\cosh|w| -1}{\cosh|w| +1} = \frac{\overline{N}_{b,\lambda}}{\overline{N}_{b,\lambda} + 2b}\label{eq:133}
  \end{align}
  and
  \begin{equation}
    \label{eq:130}
    |w| = \ln\left(\frac{1+|\lambda|}{1-|\lambda|}\right) = \ln\left[\frac{1+(1+2b/\overline{N}_{b,\lambda})^{1/2}}{(1+2b/\overline{N}_{b,\lambda})^{1/2} -1}\right]. 
  \end{equation}
  For $ \overline{N}_{b,\lambda}/2b\gg 1$ the last relation reduces to
  \begin{equation}
    \label{eq:131}
    |w| \approx \ln(2\overline{N}_{b,\lambda}/b).
  \end{equation}
  The transition probability
  \begin{equation}
    \label{eq:132}
    p_b(n \leftrightarrow \lambda) =|\langle b,n|b, \lambda\rangle|^2 = (1-|\lambda|^2)^{2b}\frac{(2b)_n}{n!}|\lambda|^{2n}
  \end{equation}
  can be expressed as (see Eq.\ \eqref{eq:133})
  \begin{align}
    \label{eq:134}
    p_b(n \leftrightarrow \lambda) = &\frac{(2b)_n(2b)^{2b}\overline{N}_{b,\lambda}^n}{n!\,(\overline{N}_{b,\lambda} +2b)^{n+2b}}, \\
     p_b(0 \leftrightarrow \lambda) = &\frac{(2b)^{2b}}{(\overline{N}_{b,\lambda} +2b)^{2b}}.\label{eq:135}
  \end{align}
  Comparing the classical quantities \eqref{eq:11} with the relations \eqref{eq:64} and \eqref{eq:120} - \eqref{eq:122} one infers the classical limit
  \begin{equation}
    \label{eq:280}
    I \asymp \lim_{(\hbar b) \to 0, |w| \to \infty} [(\hbar b)\cosh |w| \asymp  (\hbar b)\sinh |w|]
  \end{equation}
  \subsection{(Dispersive) van der Waals forces}
  F.\ London was the first one to associate the attractive van der Waals forces between neutral atoms or molecules with the nonvanishing zero-point energy
  of the HO \cite{lond1,lond2,lond3,lond4}. For atoms or molecules of the same type and without retardation  he derived - using a HO model - the potential (see also Refs.\ \cite{mil2,mil})
  \begin{equation}
    \label{eq:274}
    V(R) = -\gamma (\frac{1}{2}\hbar \omega) \,\frac{\alpha^2}{R^6},
  \end{equation}
  where $\gamma > 0$ is a number of order 1, $\alpha$ the (static) polarizability of the two atoms or molecules \cite{polar,atk3}, $R \leq \lambda = (2\pi c)/ \omega $ the
  distance of their nuclei
  and $\omega$ the angular frequency of an oscillating electric field mode which acts, e.g., either on the permanent electric dipole moments of  two molecules or
  on the their induced electric dipole moments. If $\vec{d}$ is the electric dipole moment generated at its position by the effective electric field
  $\vec{E}_{\mathit{eff}}$,
  then  $\alpha$ in an isotropic situation is defined by $\vec{d} = \alpha \, \vec{E}_{\mathit{eff}}$. It has the dimension $[L^3]$ ($\vec{E}_{\mathit{eff}}$ includes a charge factor).
  The relation \eqref{eq:274} holds for vanishing temperature $T$ (for $T > 0$ see, e.g.\ Ref.\ \cite{pass}). It is proportional to the usual HO ground state energy $\hbar \omega/2$.
  
  The potential \eqref{eq:274} is of special importance for atoms and molecules which do not have a permanent electric dipole but an induced one like, e.g. atomic hydrogen H or molecular hydrogen H$_2$, both in their ground states. The corresponding values are
  \begin{equation}
    \label{eq:282}
     \alpha[H(^1S)]= 0.67 \cdot 10^{-30} m^3,~~~ \alpha[H_2(^1\Sigma_g^+)] = 0.79 \cdot 10^{-30} m^3
  \end{equation}
 \cite{polar2}. As the polarizability $\alpha$ is closely related to the dispersion properties of an optical medium \cite{born4}, London called the forces associated
with the potential \eqref{eq:274} ``dispersion'' van der Waals forces \cite{lond4}.

If one applies London's \cite{lond2,lond3,lond4} and later heuristic arguments \cite{klep,mil2} for the derivation of the van der Waals potential to the symplectic spectrum of the HO, one obtains instead
of Eq.\ \eqref{eq:274}:
\begin{equation}
  \label{eq:275}
   V(R) = -\gamma (b\, \hbar \omega) \,\frac{\alpha^2}{R^6}.
 \end{equation} If $b< 1/2$ the ``symplectic'' van der Waals forces are weaker than the ``orthodox'' ones.

The closely related Casimir effect \cite{cas1,cas2,schw,schw2,plun,mil,most,milt,grah,gen,milt2,jaff,wre,lamo,lamb} has to be discussed separately, due to its different derivations and interpretations! 
 
  \section{Possible astrophysical  implications}
  In case the above theoretically possible ``symplectic'' - or ``fractional'' - spectra \eqref{eq:41} of the HO are - at least partially - realized in nature they could shed new light on some unsolved basic astrophysical problems of which  I shall mention the two most important ones \cite{dival}:
 
 {\it Dark energy} \cite{par1,oli,weinb,frie} and {\it dark matter} \cite{bert,bau}. Here the symplectic spectra \eqref{eq:41} may be a (the) key to the
 solutions of both problems simultaneously!

 For reasons mentioned above those spectra have not yet been seen in the laboratory. But, surprisingly, physical implications of those spectra are supported by the observationally favoured cosmological $\Lambda$CDM model \cite{peeb,ell,tur,oli,pl1} and by the associated WIMP hypothesis \cite{fen,bert,arc,schu}.
 
\emph{It probably sounds provocative, but the  observed dark energy and dark matter properties may provide the first empirical support for the existence of the  spectra \eqref{eq:41} in nature!}  

 The following discussions and arguments are mostly qualitative. The obviously necessary and crucial quantitative arguments will
 still have to be reviewed and worked out in detail!
  
  \subsection{Dark energy and the cosmological constant}
  Describing the existing astrophysical observations in terms of the  Einstein-Friedmann-Lema\^{i}tre cosmological model \cite{oli,lah} leads to the conclusion that the (``vacuum'')
  energy
  density $c^2\rho_{\Lambda}$, associated with the so-called ``Lambda''-term in the Einstein-Friedmann-Lema\^{i}tre equations, has the same order of magnitude as  the critical energy
  density \cite{lambda}
  \begin{equation}
    \label{eq:136}
    c^2\rho_{crit}= 3c^2H_0^2/(8\pi G_N) \approx 10^{-5}h^2\,\text{GeV}\,\text{cm}^{-3},
  \end{equation}
  where the scale factor $h$ for the present Hubble expansion rate $H_0$ has the approximate value $h \approx 0.7$ (this ``astrophysical'' $h$ is not to be confused with Planck's constant in the following):

  Taking into account that the observed ``dark'' energy density $c^2\rho_{\Lambda}$ is about $0.7$ of the critical density \eqref{eq:136} \cite{weinb}
  and equating $c^2\rho_{\Lambda}$
  with the vacuum energy density of the quantized free electromagnetic field \cite{kas2},
  \begin{equation}
    \label{eq:137}
    u_{em;0}(\hat{\omega},b) = \frac{b\,\hbar}{4\pi^2\,c^3}\, \hat{\omega}^4,
  \end{equation}
  where $\hat{\omega}$ is an appropriate cutoff for the corresponding divergent frequency integral 
  \begin{equation}
    \label{eq:138}
    u_{em;0}(b) = \frac{b\,\hbar}{\pi^2\,c^3}\,\int_{\omega \geq 0} d\omega\, \omega^3,
  \end{equation}
  allows to make a numerical estimate of $b$:

  Introducing the cutoff length
  \begin{equation}
    \label{eq:139}
    \ell = \frac{2\pi c}{\hat{\omega}}
  \end{equation} leads to the approximate equality
  \begin{equation}
    \label{eq:140}
     u_{em;0}(\ell,b) = b \frac{4\pi^2\,\hbar c}{ \ell^4}\, \approx 0.7 \cdot c^2\rho_{crit}.
   \end{equation}
   Taking for $\ell$ the (reduced) Compton wave length of the electron,
   \begin{equation}
     \label{eq:141}
 \ell \approx   \lambdabar = \hbar/(m_ec) \approx 3.9\cdot 10^{-11}\text{cm}  
   \end{equation}
   and inserting $h^2 \approx 0.5$ into relation \eqref{eq:136}
   gives for the B-index $b$ the approximate value
   \begin{equation}
     \label{eq:142}
     b \approx 10^{-35}\,.
   \end{equation}
   This is an extremely small $b$ -- value, but it is  theoretically allowed in the present framework! This is in contrast to the conventional theoretical
   estimates of the dark energy with $b = 1/2$ \cite{wein,car1,car2,frie} which represent the most embarrassing discrepancy between observations and theoretical reasoning
   in all of present-day physics!

   The discussion above assumes that all modes have the same index $b$. This simplification is, of course, not necessary. $b$ can depend on the frequency $\omega$:
   $b=b(\omega) \in (0,1]$. It can, therefore, become a dynamical quantity!

   The estimate \eqref{eq:142} depends sensitively on the choice of the cutoff length $\ell$: if we. e.g., replace the factor $10^{-11}$ by $10^{-10}$ the estimate in
   Eq.\ \eqref{eq:142} is reduced to $ b \approx 10^{-31}$. In addition all non-electromagnetic effects were neglected (they would lead to an even smaller value of $b$ than that in Eq.\ \eqref{eq:142}!). This will be justified by the discussion below
   concerning the nature of dark matter as being essentially molecular hydrogen the dynamics of which is essentially determined by electromagnetic forces.

   Let me make another very crude estimate related to the ``cosmic'' order of magnitude \eqref{eq:142} of the B-index $b$: Consider the relation \eqref{eq:49} between 
   the angular frequency $\omega = 2\pi \nu$, the time period $T_{2\pi;b}$ and the index $b$: Most of the very first molecules and molecular ions  after the beginning of the recombination epoch in the very early universe  were diatomic,
   with the vibrating nuclei locally  emitting infrared light with frequencies $\nu =\omega/(2\pi)$ around $10^{14}~$s$^{-1}$. Even though
   homonuclear
   elements like $H_2$ do not have a permanent electric dipole element, they still radiate in the infrared \cite{rou} and especially can emit Raman radiation
  by induced dipole momennts (see Ch.\ IV.B above).

   Taking for $T_{2\pi;b}$ the extreme value $10^{10}$ yr $ \approx 3\,\cdot 10^{17}$~s and ignoring cosmic red shifts $z$ (i.e.\ being in the rest frame of the molecule)
   we get the crude estimate
   \begin{equation}
     \label{eq:144}
     b = 1/(\nu\,T_{2\pi;b}) \approx 10^{-32}.
   \end{equation}
   An important open question is whether the vacuum (``dark'') energy has changed with cosmic time which would imply a corresponding time dependence of $b\,$!
   \subsection{``Dark'' b-H$_2$ and other primordial molecules \\ as dark matter?}
   The following most intriguing but perhaps also dangerously seductive or even deceptive attempt intends to interpret the cosmic ``dark matter'' in the ``symplectic'' framework
   of the HO. The central hypothetical role here is being played by ``symplectic'' molecular hydrogen b-H$_2$ as the main candidate for dark matter. The possibility that molecular hydrogen H$_2$ may play a role for the understanding of dark matter has been tentatively discussed before \cite{carr,pfe1,pfe2,pfe3,comb1,comb2,pao1,pao2,pao3,pao4,silk} without, it seems, having a lasting impact. But the possible existence of a (weak or hidden) ``detuned'' symplectic spectrum of the vibrating b-H$_2$ allows for a new and probably more promising approach! In addition to this "symplectic" detuning there is, of course, the usual cosmic redshift $z$ due to the expansion of the universe.

   All the directly obtained  experimental and obervational data like those of the ``Planck'' Kollaboration etc.
   are, of course, not affected, but all the {\it calculated  particle and cosmic standard model} dependent dynamical - vibration related - electromagnetic properties (transition probabilities of emissions, absorptions, dissociations, ionizations and other
   rates etc.) have to be re-evaluated. The same applies to the Big Bang Nucleosynthesis and  the primordial photon-baryon ratio \cite{cyb,fiel}!

   Molecular hydrogen plays already an important role in the present standard (``orthodox'') cosmological paradigm \cite{field,peeb,will,comb1,gal1,lepp,snow,gal,loeb,uba,wak}. Due to the missing electric dipole moment it is difficult to detect astrophysically. For searches in the intergalactic medium (IGM) one uses a
   plausible correlation between the densities of H$_2$ and of carbon monoxyde CO \cite{bol}, which has a permanent electric dipole moment and is more visible. But the atoms C and O
   are not primordial ones and have to be bred in (first) stars etc..

   Presently, however, we are primarily interested in the epoch of the universe which is called its ``Dark Ages'' \cite{loeb0,mir,loeb1,loeb2,koop}, i.e.\ the cosmic time period which started
    when the photons decoupled from matter and  primordial atoms (mainly He and H) could form (``recombination epoch'' \cite{peeb1,pea,pea1,loeb0,mukh,mukh1,wein2,wein3}) at about 400000 years after the big bang (at redshift $z = z_{\star} \approx 1100 \approx 3000^{\circ}$\,K$ \approx 0.26$ eV \cite{pl6}). And which ended
   just before (around $z\approx 30$, i.e.\ about 80 Myr after the big bang) density fluctuations of the primordial gases led to the first gravitational ``clumps'' as
   seeds for the first stars \cite{loeb0,naoz,loeb,fial} and the first galaxies \cite{bro,loeb}. The heat and radiation associated whith this gravitational process reionized the primordial neutral gases of the dark ages \cite{fan,loeb}, a cosmic period  called the ``Dawn'' of the universe \cite{koop}.

   During those ``dark ages'' there were no ``dust grains'', no ``metals'' (like $^{16}$O or $^{12}$C etc.) and no X- and no cosmic rays. There were only gases of primordial photons, neutrinos, electrons, protons, deuterons and atoms He, H, D, Li, their ions
   and first primordial diatomic molecules and their ions \cite{gal1,lepp,gal} like H$_2$ and, e.g.\ (HeH)$^+$ \cite{dab,zyg,bov,gal,gue}  most of them at least partially in local thermodynamical equilibrium. Ions may be ``thrown'' out of equilibrium by primordial magnetic fields.

   As the main observable signal from that epoch the 21~cm line from the electron spin flip in the field of the magnetic moment of the proton in atomic hydrogen has intensively been discussed more recently \cite{loeb0,fur,prit,bark2,cohe,bark3,novo2,koop} and experimentally implemented or planned \cite{mon,bow,koop}. Molecular hydrogen is also assumed to play an essential role during the dark ages \cite{loeb0,hira,ali,novo1} because of its  cooling properties \cite{loeb,bark2}.

   In the following a number of possibly important properties of ``symplectic'' molecular hydrogen, called b-H$_2$,  are listed which support the hypothesis of essentially identifying dark matter with b-H$_2$. The $b$-detuned spectra of other primordial diatomic molecules (containing D, He and Li \cite{gal1,lepp,gal}) probably play an essential role, too, but that of b-H$_2$  is the most important one and here serves as a prototype for a symplectically detuned molecular HO! The following discussion relies on properties of b-H$_2$ pointed out in Ch.\ IV.B above. 
   
What is still missing in the discussion presented here is a necessary critical \emph{quantitative} evaluation! This is, of course,  due to the fact that up to now no corresponding  experimental laboratory data are   available and - above all - that the associated theory for the dynamics of the ``symplectic'' spectroscopy still has to be worked out! But in principle it should be possible to decide the empirical
existence of those symplectic spectra in the laboratory, keeping in mind that the  index $b$ might be very small (see Eq.\ \eqref{eq:142}) and, therefore, optical transitions may be very weak (slow)!

Here is a list of properties and problems which shows how the b-H$_2$ (plus other primordial molecules) dark matter hypothesis compares with  what essentially is known about dark matter experimentally:
   \begin{enumerate} \item It is quite surprising that the measured (relative) cosmic baryonic density $\Omega_b \approx 0.05$ and the dark matter density $\Omega_d \approx 0.25$ are roughly of the same order of magnitude. And that the total matter density $\Omega_m \approx 0.3$ is about of the same order of magnitude as the
     dark (``vacuum'') energy density $\Omega_{\Lambda} \approx 0.7$! But this is no longer surprising if dark matter consists essentially of ``b-''H$_2$ and other
     ``b''-detuned primordial molecules, i.e.\ it is
     baryonic, too\,! In addition there may be some kind of  dynamical relationships between vacuum energy and matter.   \item Whereas there appears to be - on average -  5 times more dark than baryonic matter in the universe, not a single non-standard particle has been detected, despite years-long tremendous and ingenious efforts by  experimentalists and theoreticians \cite{fen,bert,baud,arc,sal,schu}. That failure is no surprise if dark matter essentially consists of b-H$_2$ and other
     b-detuned primordial diatomic molecules and ions! \item If indeed dark matter consists of such molecules and ions, this implies that there is no dark matter in the universe before the recombination/decoupling era  when the formation of
          primordial atoms (He, H, D, Li), their molecules and ions sets in \cite{gal1,lepp,loeb,gal}. And there is then no annihilation of dark matter! \item  For the formation of b-H$_2$ etc.\ as dark matter about 5 times more primordial baryons and electrons are required. This  needs, of course, a re-evaluation of the orthodox big bang nucleosynthesis (BBN) and the related baryon to photon ratio \cite{pea,mukh,wein2,cyb,fiel,oli}. \item The dark matter approach discussed here may also shed some new light on the ``Lithium problem'' of the current BBN interpretation (a disturbing discrepancy between the observed Li/H ratio und the BBN predicted one) \cite{lith}. \item If there are at  the beginning of the recombination phase - ignoring all other light primordial nuclei, their atoms, molecules and ions -  protons and  electrons, why should they form dominantly  b-H$_2$ molecules  and  H atoms so that on average we have a ratio of approximately 5:1? A preliminary argument is that the b-H$_2$ molecules are energetically favoured compared to 2 free H atoms because of their binding energies (4.48 +
     0.26 = 4,74)\,eV (see Eqs.\ \eqref{eq:249} and \eqref{eq:248}). In order to understand the observed ratio 5:1 of ``dark'' matter to baryonic matter one probably
     has to take
     the dynamics of the different atomic and molecular reactions  into account! Note that the formation of primordial hydrogen molecules can take different
     routes \cite{gal1,lepp,gal}. \item It  also is worth mentioning that the ``partial'' Boltzmann factor \eqref{eq:255} becomes smaller than of order 1 just at the beginning of the dark ages. This implies that as lower temperatures statistically the lower b-zero-point energy levels are  correspondingly more populated than the ``orthodox'' ones!
   \item  Even if the transitions between the detuned b-levels of the vibrating molecules are very weak there should be enough time (60-100 Myr) available during the dark ages to ``settle
     down'' to the b-zero-point energy levels.
         \item Galaxies (first small ones which then became larger)
           are assumed to have grown from (Jeans) density perturbations/fluctuations leading to instabilities inside  halos of dark matter \cite{loeb,wech,sal,zan,ina}. That appears quite ``natural''   if those halos are essentially cold remains of primordial diatomic molecular (mainly hydrogen) clouds from the dark ages, in which most of the
           molecules are not yet dissociated. Consequently one  has to expect  dynamical relations between the ionized core of galaxies (including their stars, black holes and gases) and their dark cold molecular halos.
    \item If dark matter consists essentially of b-H$_2$ and other primordial diatomic molecules and ions then there was a rich amount of matter for the formation of baryonic supermassive black holes in the center of evolving galaxies \cite{brom2,ina}. \item As H$_2$ is considered to be an important coolant for the primordial gases before the formation of cosmic structures sets in \cite{bour,cop,loeb} the cooling by b-H$_2$ has to be analyzed anew.
    \item Molecular hydrogen has a mass of about 2 GeV. Comparing this with the reconstructed temperature $T_{\gamma} \approx 0.26$ eV of the photon gas at the  beginning of the recombination period and assuming the matter gases to have roughly the same temperature about that time implies that the molecules are highly nonrelativistic. The same applies, of course, to the other primordial molecules and ions containing He, D, Li etc. \item Neutral H$_2$ molecules interact weakly among themselves and with other neutral atoms or molecules by van der Waals forces \cite{mck,flow4}. In addition, the relation \eqref{eq:275} shows that the strength  of van der Waals forces becomes weaker than the ``orthodox'' ones for $b< 1/2$. For the interactions of ``orthodox'' H$_2$ molecules  with atomic hydrogen and helium see, e.g., the Refs.\
      \cite{truh,flow1,flow3,flow2}. \item A theoretical
      analysis
      of the data obtained by the 21-cm radio wave detector EDGES \cite{bow} concluded \cite{bark,fial2,fial1,coh3} that most likely there are interactions beyond the gravitational ones between the primordial
      atomic hydrogen and the dark matter particles, the latter having a mass of about  a few GeV! All this fits - at least qualitatively - the above b-H$_2$
      interpretation of dark matter very well. In addition it supports the hypothesis that  dark matter consists of (w)eakly (i)nteracting (m)assive (p)articles
      (``WIMPs'').
      \item The recently observed discrepancy between gravitational lensing \cite{mene} and computer-simulated (standard) dark matter models may also find an explanation within the framework discussed in the present paper! 
      \item Of special interest in the context of the near-infrared ``symplectic'' spectra of H$_2$ etc.\ molecules discussed above are the recent unaccounted cosmic optical background observations in  that spectral region by the CIBER and New Horizons collaborations \cite{mats,lauer}.
      \end{enumerate} All the properties listed above are - qualitatively - surprisingly  compatible with the current $\Lambda$CDM model of the universe and the WIMP
      hypothesis for dark matter!
      
      All this suggests that the hypothesis: ``the dark matter observed in the universe consists essentially of b-H$_2$ and of a smaller amount of other primordial diatomic b-detuned molecules and their ions''  should be taken seriously and analyzed more quantitatively and experimentally as well. This requires  joint efforts of the physisists involved in the field!

      One particular lesson to be learnt from the discusssions above is that we  do not know yet enough  about the physics of the  cosmic quantum``vacuum''!
      
\begin{acknowledgments} I very much thank the DESY Theory Group for its enduring  kind and supportive hospitality
after my retirement from the Institute for Theoretical Physics of the RWTH Aachen. I am grateful  to David
Kastrup for providing the figures and for further technical support. \end{acknowledgments}

   \begin{appendix}
     \section{Hilbert space for $\tl{K}_j$ and $\tl{C}_g(K)$ on $\mathbb{R}_0^+$ }
     \subsection{Representation of the Lie algebra generators $\tl{K}_j$ on $L^2(du,\mathbb{R}^+_0)$}
     The Hilbert spaces of square-integrable functions 
\begin{equation}
  \label{eq:77}  
  f(u) = u^{\alpha/2}\,e^{-u/2}g(u),\; u \geq 0, \, \alpha > -1,
\end{equation}
with the scalar product 
\begin{align}
  \label{eq:78}
  (f_2,f_1) \equiv & \int_0^{\infty}du\,f_2^{\ast}(u)f_1(u)\\ =& \int_0^{\infty}du\,u^{\alpha}e^{-u}\,g_2^{\ast}(u)\,g_1(u). \nonumber
\end{align}
also provide  Hilbert spaces for irreducible unitary representations of the group $SO^{\uparrow}(1,2)$, its twofold symplectic covering group $Sp(2,\mathbb{R})$ and all
other covering groups \cite{wolf,ka13}.

The usual orthonormal basis for a fixed $\alpha$ is given by the associated Laguerre functions
\begin{align}
  \label{eq:79}
  \hat{e}_{\alpha;n}(u) =& \sqrt{\frac{n!}{\Gamma(n+ \alpha +1)}}\; u^{\alpha/2}\,e^{-u/2}\,L^{\alpha}_n(u),\,n=0,1,2, \ldots \\
                       & (\hat{e}_{\alpha;m},\hat{e}_{\alpha;n}) = \delta_{mn}\,, \nonumber
 \end{align}
where the associated Laguerre polynomials $L^{\alpha}_n(u)$ are defined as
\begin{align}.
  L^{\alpha}_n(u)=& \sum_{m=0}^{m=n}{n+\alpha\choose n-m} \frac{(-u)^m}{m!},\; 
    L^{\alpha}_n(0)= \frac{(\alpha + 1)_n}{n!}, \\ & (a)_n = a\,(a+1)\,(a+2)\ldots \,(a+n-1). \nonumber
\end{align}
Examples:
\begin{equation}
  \label{eq:94}
  L^{\alpha}_0(u) = 1,~~L^{\alpha}_1(u) = \alpha+1-u.
\end{equation}
The polynomials $ L^{\alpha}_n(u)$ obey the differential equation
\begin{equation}
  \label{eq:80}
  u\,\frac{d^2L^{\alpha}_n}{du^2} + (\alpha -u +1)\frac{dL^{\alpha}_n}{du} + n\,L^{\alpha}_n =0.
\end{equation}
The operators $\tl{K}_j$ here have the explicit form \cite{ka0,ka8}, with $\alpha = 2b-1$,
\begin{align}
  \tl{K}_0 =& -u\,\frac{d^2}{du^2}\,-\frac{d}{du} + \frac{(2b-1)^2}{4u} + \frac{u}{4}, \label{eq:81}\\
  \tl{K}_1 =& -u\,\frac{d^2}{du^2}\,-\frac{d}{du} + \frac{(2b-1)^2}{4u} - \frac{u}{4}, \label{eq:82}\\
  \tl{K}_2 =& \frac{1}{i}(u\,\frac{d}{du} + \frac{1}{2})\,. \label{eq:83}
\end{align} They obey the Lie algebra \eqref{eq:32}. The inequality $\alpha > -1$ (Eq.\ \eqref{eq:77}) implies $b>0$, as desired!

Eigenfunctions $f_{b;n}(u)$ of $\tl{K}_0$ are the basis functions \eqref{eq:79} with $\alpha = 2b-1$:
\begin{align}
  \tl{K}_0f_{b;n}(u) =& (n+b)\,f_{b;n}(u),\,n= 0,1,2,\ldots \label{eq:84} \\
  f_{b;n}(u) =& \hat{e}_{2b-1;n}(u)\nonumber \\ =& \sqrt{\frac{n!}{\Gamma(2b+n)}}\; u^{b-1/2}\,e^{-u/2}\,L^{2b-1}_n(u). \nonumber
\end{align}
The ground state function is
\begin{equation}
  \label{eq:95}
  f_{b;0}(u) = \frac{1}{\sqrt{\Gamma(2b)}}\,u^{b-1/2}\,e^{-u/2}.
\end{equation}
It obeys 
\begin{equation}
  \label{eq:143}
  K_-f_{b;0}(u) = (K_1 - iK_2)f_{b;0}(u) = 0.
\end{equation}
The expressions \eqref{eq:81} and \eqref{eq:82} allow for an interpretation of the variable $u$ in terms of the classical variables
$\vp$ and $I$: We have
\begin{equation}
  \label{eq:85}
  \tl{K}_0-\tl{K}_1 = \frac{u}{2}.
\end{equation}
Comparing this with the difference
\begin{equation}
  \label{eq:86}
  h_0-h_1 = v \equiv I(1-\cos\vp) = 2I\,\sin^2(\vp/2)
\end{equation} we note the correspondence
\begin{equation}
  \label{eq:87}
  u \leftrightarrow 2v =4I\,
  \sin^2(\vp/2).
\end{equation}
The relation \eqref{eq:86} can be interpreted geometrically as follows:

$I > 0$ and $\vp \in [0, 2\pi)$ can be considered as polar coordinates of a plane where $\vp =0$ describes the positive part of the abscissa. According to Eq.\
\eqref{eq:86} the variable $v$ can be considered as the
difference between the distance $I$ of the point $(\vp=0,I)$ on the abscissa and the projection of
the position vector $(\vp,I)$ on the abscissa. From Eq.\ \eqref{eq:87} we have
\begin{equation}
  \label{eq:252}
  v_1 \in [0, 2I] \text{ for } \vp \in [0, \pi)],\, v_1 \in \mathbb{R}^+_0
\end{equation}
and
\begin{equation}
  \label{eq:253}
  v_2 \in [2 I, 0] \text{ for } \vp \in [\pi, 2\pi], \, v_2 \in \mathbb{R}^+_0.
\end{equation}

Time reflection
\begin{equation}
  \label{eq:276}
  T:~~\vp \to -\vp \text{ plus complex conjugation },
\end{equation}
can be implemented as expected:
\begin{equation}
  \label{eq:277}
  T:~~ u \to u,~ \tl{K}_0 \to \tl{K}_0,~ \tl{K}_1 \to \tl{K}_1,~\tl{K}_2 \to - \tl{K}_2
\end{equation}
The homeomorphic doublings \eqref{eq:252} plus \eqref{eq:253} may be important for the implementation of  space reflections (see Ch.\ III.E above)
\begin{align}
  \label{eq:254}
  P:~~&  \vp \to \vp + \pi,~~ I \to I, ~~~ v_1 \leftrightarrow v_2, \\
 & \tl{K}_0 \to \tl{K}_0,~ \tl{K}_1 \to - \tl{K}_1,~\tl{K}_2 \to - \tl{K}_2. \nonumber
\end{align}
So,  in order to implement space reflections, it appears one has to double the Hilbert space defined by Eqs.\ \eqref{eq:77} and \eqref{eq:78}.

\subsection{Eigenfunctions and spectrum of $\tl{C}_g = \tl{K}_0 +g\,\tl{K}_1$}
The eigenfunctions $f_{g,b;n}$ of the operator $\tl{C}_g = \tl{K}_0 +g\,\tl{K}_1$ can be obtained from the eigenvalue Eq.\ \eqref{eq:84}:

  We have
  \begin{align}
    \label{eq:88}
   & \tl{C}_g(u)f_{g,b;n}(u) = (\tl{K}_0 + g\,\tl{K}_1)(u)f_{g,b;n}(u)  \\ &= (1+g)\biggl[ -u\,\frac{d^2}{du^2}\,-\frac{d}{du} + \frac{(2b-1)^2}{4u} \biggr] f_{g,b;n}(u) \nonumber \\
    & + (1-g)\frac{u}{4}f_{g,b;n}(u)= \tl{c}_{g,b;n}f_{g,b;n}(u). \nonumber
  \end{align}
  With
  \begin{equation}
    \label{eq:89}
    u = \sqrt{\frac{1+g}{1-g}}\,v,
  \end{equation}
  
  we get
  \begin{equation}
    \label{eq:90}
    \tl{C}_g(u) = \sqrt{1-g^2}\tl{K}_0(v).
\end{equation}
Therefore $\tl{C}_g(u)$ has the eigenvalues
\begin{equation}
  \label{eq:91}
  \tl{c}_{g,b;n}= (n+b)\sqrt{1-g^2},\,n=0,1,2,\ldots
\end{equation}
and the eigenfunctions
\begin{equation}
  \label{eq:92}
  f_{g,b;n}(u) = f_{b;n}[v(g,u)].
\end{equation}
Thus, the Hamiltonian
\begin{equation}
  \label{eq:96}
  H=\hbar\omega\,\tl{C}_g
\end{equation}
has the eigenvalues
\begin{equation}
  \label{eq:97}
  \hbar\omega_g(n+b),\; \omega_g = \sqrt{1-g^2}\,\omega,
\end{equation}
where $\omega_g$ is the effective frequency \eqref{eq:93} of the classical system and the result \eqref{eq:97} coincides with the one obtained in Ch.\ III.E.5 above.
  \section{Notational conventions for the states of diatomic molecules}
  Here the essential features of the conventional notations for the states of diatomic molecules are briefly recalled \cite{herzb1,land2,nist1}, for a better understanding of the
  discussion in Ch.\ IV.B above:
  
  Starting point for the description employed is the  BO approximation \cite{herzb2} which makes use of the small ratio of electron and nucleon masses:
  This leads to a separation of the original Schr\"{o}dinger equation into two simpler ones: one for the electrons in their mutual Coulomb potentials and in those of
  the nuclei held fixed,  and second one for the two nuclei moving in  potentials provided by
    solutions $E_{el}(R.a)$ of the electronic Schr\"{o}dinger equation which depend ``adiabatically'' on the internuclear distance $R =|\vec{R}_2 - \vec{R}_1|$ and  possibly on
  other parameters $a$ of the nuclei etc.  Qualitatively it is important to differentiate between homonuclear molecules (same nuclei) and
  heteronuclear ones (different nuclei). In the homonuclear case there is no permanent electric dipole moment and, therefore, no corresponding light emission or absorption!

  The main coordinate refererence is the internuclear axis (INA), i.e.\ the straight line passing through the two nuclei. It is essential for the description of electronic motions and the
  nuclear ones as well. The following remarks apply to diatomic molecules only, including their ions.
  \subsection{Electronic motions}
  Like in the case of atoms one starts by ignoring spin-orbit couplings ($\hbar = 1$ in the following): the absolute values $\Lambda = |M_L|= 0,1,\ldots L$ of the projections of the total electronic orbital angular momentum $\vec{L}$ on the INA are denoted by $\Sigma\, (\text{for } \Lambda = 0), \Pi \,(\text{for } \Lambda = 1),
  \Delta\, (\text{for }\Lambda = 2), \ldots$ (in analogy to the atomic $s\, (\text{for } l=0), p\, (\text{for } l=1), d\, (\text{for } l=2), \ldots)$, where the state $\Sigma$ is non-degenerate and the states $\Pi, \Delta, \ldots$ are 2-fold degenerate.

  The following symmetries further specify  a state:

Reflections on any plane containing the INA form a symmetry of diatomic molecules. For non-degenerate states like $\Sigma$ this means that its state vector stays invariant or changes
sign. Thus one can have $\Sigma^+$ or $\Sigma^-$. For degenerate states like $\Pi, \Delta, \ldots$ a 2-dimensional state vector may be mapped into another one and no
definite ``parity'' can be assigned. This holds for homo- and heteronuclear diatomic molecules alike.

For homonuclear diatomic molecules - like H$_2$ - there is still another reflection symmetry \cite{wign}: reflections on the midpoint of the INA between the two nuclei transforms the molecule onto itself. Correspondingly the wave function may be even (= $g$ from German ``(g)erade'') or odd (= $u$ from German ``(u)ngerade''). This holds also for
$\Lambda \neq 0$. So one can have the electronic states $\Sigma^{\pm}_g,\Sigma^{\pm}_u, \Pi_g, \Pi_u, \ldots$.

For electric dipole transitions between different electronic levels the following selection rules hold \begin{center}
$  +\leftrightarrow -,~ + \not \leftrightarrow +,~ -  \not \leftrightarrow -\,; ~ g \leftrightarrow u,~ g \not \leftrightarrow g,~ u
\not \leftrightarrow u.  $ \end{center}

Denoting the projection of the total electron spin on the INA by $S$ one can have $2S+1$ multiplets, i.e.\ singlets and triplets for diatomic molecules. The corresponding states are denoted as usual, e.g.\ $^{2S+1}\Sigma^+_g$ etc.

Different electronic levels may have the same $^{2S+1}\Lambda^{\pm}_{g,u}$. If they are singlets ($S=0$) one differentiates between them by the capital letters X (ground state),  A, B, $ \ldots$. If they are non-singlets one writes a,b, $\ldots$. For the electronic ground state (potential) of H$_2$ one has X $\,^1\Sigma_g^+$ (see Fig.\ 4). The choice of the capital letter for different electronic states is
not stringent, but can have historical backgrounds.
\subsection{Nuclear motions}
As mentioned before, the electronic energy states $E(R,a)$ just discussed depend on the distance $R$ of the two nuclei and   serve as  potentials for their motions, as indicated in Fig.\ 4 and  discussed in more detail in the literature \cite{bofig}. Again, the vibrational energy levels in those potentials are denoted by $v = 0,1, \ldots$, where the lowest levels are approximately ``harmonic''. In addition the molecule can rotate around any axis passing through the center of mass on the INA and being orthogonal to the latter. The associated rotational quantum number is generally denoted by $J$.

If electronic configurations $E(R,a)$ have (local) minima, these generally lead to several vibrational states and, therefore, one can have  transitions (emissions and absorptions) ($v'' \leftrightarrow v'$)  between vibrational levels of different electronic states. This leads to so-called ``bands'' \cite{herzb4}. In addition, each vibrational state ($v$''  and $v$') can be associated with several rotational states ($J$'' and $J$') leading to different ``branches'' \cite{herzb5}.
  \end{appendix}
\bibliography{wignerfunction}

\begin{thebibliography}{240}
\bibitem [{kas({\natexlab{a}})}]{kas}%
  \BibitemOpen
  \href@noop {} {}\bibinfo {note} {{A}nother frequent choice is $ q
  =\sqrt{2I/m\,\omega} \sin \vp,\, p =\sqrt{2m\,\omega\,I} \cos
  \phi$.}\BibitemShut {Stop}%
\bibitem [{ka1({\natexlab{a}})}]{ka10}%
  \BibitemOpen
  \href@noop {} {}\bibinfo {note} {{D}etailed arguments can be found in
  {A}ppendix {A} of Ref.\ \cite{ka7}, {C}h.\ 4 of Ref.\ \cite{ka8} and {C}h.\ 2
  of Ref.\ \cite{ka9}}\BibitemShut {NoStop}%
\bibitem [{ka1({\natexlab{b}})}]{ka11}%
  \BibitemOpen
  \href@noop {} {}\bibinfo {note} {The model has been mentioned in Refs.\
  \cite{ka8} and \cite{ka9}}\BibitemShut {NoStop}%
\bibitem [{kas({\natexlab{g}})}]{kas9}%
  \BibitemOpen
  \href@noop {} {}\bibinfo {note} {For work on anyons see {C}h.\ I in Ref.\
  \cite{ka} with the Refs.\ [14]-[25] . More recent papers on anyons are
  Refs.\, \cite{rose} and \cite{bart}}\BibitemShut {NoStop}%
\bibitem [{mil()}]{mil2}%
  \BibitemOpen
  \href@noop {} {}\bibinfo {note} {For a more advanced discussion of the
  relation between van der {W}aals forces and electromgagnetic vacuum energies
  see, e.g., Ref.\ \cite{mil}, {C}hs.\ 3.11, 7 and 8. A quote: "Thus we can
  derive both the retarded and unretarded van der Waals interactions between
  two molecules from the $macroscopic$ theory of the force between two
  dielectric plates. This theory requires basically only the Maxwell equations
  and the assumption that each mode of the field has a zero-point energy $\hbar
  \omega/2$" (p.\ 232)}\BibitemShut {NoStop}%
\bibitem [{atk()}]{atk3}%
  \BibitemOpen
  \href@noop {} {}\bibinfo {note} {See Ref.\ \cite{atk2}, {C}h.\ 12; the
  authors here appear as not having known London's paper Ref.\ \cite{lond3}
  where he clearly discussed and emphasized the ground state energies, the
  ionization ones only as a rough approximation in Ref.\
  \cite{lond4}}\BibitemShut {NoStop}%
\bibitem [{bof()}]{bofig}%
  \BibitemOpen
  \href@noop {} {}\bibinfo {note} {See, e.g. the Figs.\ 2 and 3 in Ref.\
  \cite{sha}, Fig.\ 3 on page 2.11 of Ref.\ \cite{will}, Fig.\ 3 in Ref.\
  \cite{wak} and Fig. 2 in Ref.\ \cite{bail}}\BibitemShut {NoStop}%
\end{thebibliography}
\end{document}